\documentclass[conference]{IEEEtran}
\IEEEoverridecommandlockouts
\usepackage{cite}
\usepackage{algorithmic}
\usepackage{graphicx}
\usepackage{xcolor}

\usepackage{amssymb,amsmath,amsthm, amsfonts}
\usepackage{mathrsfs}
\usepackage{subfigure}
\usepackage{caption}
\usepackage{float}
\usepackage{booktabs}
\usepackage{flushend}
\theoremstyle{plain}
\newtheorem{theorem}{Theorem}
\newtheorem{lemma}{Lemma}
\newtheorem{problem}{Problem}
\newtheorem*{remark}{Remark}
\theoremstyle{definition}
\newtheorem{example}{Example}

\begin{document}

\title{Revisiting RFID Missing Tag Identification
}

\author{\IEEEauthorblockN{Kanghuai Liu\IEEEauthorrefmark{1}, Lin Chen\IEEEauthorrefmark{1}\IEEEauthorrefmark{2},~\IEEEmembership{Member,~IEEE}, Jihong Yu\IEEEauthorrefmark{3},~\IEEEmembership{Member,~IEEE}, Junyi Huang\IEEEauthorrefmark{1}, and Shiyuan Liu\IEEEauthorrefmark{1}}
\IEEEauthorblockA{\IEEEauthorrefmark{1}School of Computer Science and Engineering, Sun Yat-sen University, Guangzhou, China}
\IEEEauthorblockA{\IEEEauthorrefmark{2}Key Laboratory of Machine Intelligence and Advanced Computing, Ministry of Education, China}
\IEEEauthorblockA{\IEEEauthorrefmark{3}School of Information and Electronics, Beijing Institute of Technology, Beijing, China}
\IEEEauthorblockA{Email: \{liukh8@mail2, chenlin69@mail, huangjy233@mail2, liushy79@mail2\}.sysu.edu.cn, jihong.yu@bit.edu.cn}}


\maketitle

\begin{abstract}
We revisit the problem of missing tag identification in RFID networks by making three contributions. Firstly, we quantitatively compare and gauge the existing propositions spanning over a decade on missing tag identification. We show that the expected execution time of the best solution in the literature is $\Theta \left(N+\frac{(1-\alpha)^2(1-\delta)^2}{ \epsilon^2}\right)$, where $\delta$ and $\epsilon$ are parameters quantifying the required identification accuracy, $N$ denotes the number of tags in the system, among which $\alpha N$ tags are missing. Secondly, we analytically establish the expected execution time lower-bound for any missing tag identification algorithm as $\Theta\left(\frac{N}{\log N}+\frac{(1-\delta)^2(1-\alpha)^2}{\epsilon^2 \log \frac{(1-\delta)(1-\alpha)}{\epsilon}}\right)$, thus giving the theoretical performance limit. Thirdly, we develop a novel missing tag identification algorithm
by leveraging a tree structure with the expected execution time of $\Theta \left(\frac{\log\log N}{\log N}N+\frac{(1-\alpha)^2(1-\delta)^2}{ \epsilon^2}\right)$, reducing the time overhead by a factor of up to $\log N$ over the best algorithm in the literature. The key technicality in our design is a novel data structure termed as collision-partition tree (CPT), built on a subset of bits in tag pseudo-IDs, leading to more balanced tree structure and reducing the time complexity in parsing the entire tree.
\end{abstract}


\section{Introduction}

Characterized by backscatter communication, fast and accurate identification, and non-line-of-sight tracking capability, radio frequency identification (RFID) has
become an enabling technology in emerging IoT applications and has been widely applied in a large number of applications ranging from warehouse management, supply-chain control, indoor localization, to object tracking~\cite{ZhuW_2020,Jihong_2020,LiuX_2014}. 
An RFID system in its standard form is composed of a reader, usually connected to a back-end server, and a number of tags that can harvest energy from the reader via RF waves to communicate with it over the wireless medium~\cite{ChiuC_2008}. 
 
In typical large-scale RFID applications such as inventory control and factory surveillance, one of the most important and fundamental tasks is missing tag identification. This task is particularly challenging, because it needs to be performed in a time-efficient fashion while satisfying the identification accuracy requirement, two objectives that are often at odd with each other~\cite{Shahzad_2016,JSu_2020}. 
Given its paramount practical importance, missing tag identification has been extensively investigated,
with the first proposition dating back to as early as 2010~\cite{LiT_2010}. Since then, a variety of techniques involving adaptive sampling~\cite{LiT_2013}, probabilistic framed ALOHA~\cite{LiuX_2014}, multi-seed hashing~\cite{LiuX_2015}, physical layer coding~\cite{ZhengY_2013} etc., have been delicately mobilized to design dozens of missing tag identification algorithms, some focusing on particular application scenarios, some searching for appropriate performance trade-offs among detection accuracy, time and energy cost, others advancing the state of the art by refining the related algorithm and protocol design so as to reduce the communication and computing overhead. Given the large body of research spanning over a decade~\cite{LiT_2010, LiT_2013, ZhengY_2013, LiuX_2014, LiuX_2015, ShaoC_2015, ZhangL_2017} resulting in a large palette of proposed solutions, two natural questions arise:
\begin{enumerate}
    \item How do they compare with each other? Can we quantify their performance?
    \item Is there a theoretical limit for any missing tag identification algorithm? If yes, is it achievable?
\end{enumerate}

Driven by the above research questions, and standing on the shoulder of the existing works, in this paper we revisit the problem of missing tag identification both retrospectively and prospectively. 

Retrospectively, we quantitatively compare and gauge the existing propositions spanning over a decade on missing tag identification (cf. Tab.~\ref{tab:comparison}). Specifically, we show that the expected execution time of the best solution in the literature is $\Theta \left(N+\frac{(1-\alpha)^2(1-\delta)^2}{ \epsilon^2}\right)$, where $\delta$ and $\epsilon$ are parameters quantifying the required detection accuracy, $N$ denotes the number of tags in the system, among which $\alpha N$ of them are missing. For comparison, we analytically establish the expected execution time lower-bound for \textit{any} missing tag identification algorithm in Theorem~\ref{thm:bound} as $\Theta\left(\frac{N}{\log N}+\frac{(1-\delta)^2(1-\alpha)^2}{\epsilon^2 \log \frac{(1-\delta)(1-\alpha)}{\epsilon}}\right)$.\footnote{Throughout this paper, logarithm is taken on base $2$.} We note that, despite the large body of research works, the theoretical performance limit of missing tag identification has not been formally characterized. Our result derived in this paper fills this important gap, and can serve as a design guideline for future research in this field. 

Prospectively, armed with the theoretical results we derive, we further develop a novel missing tag identification algorithm
by leveraging a tree-based structure to optimize the time efficiency, the central performance metric in missing tag identification. With the expected execution time of $\Theta \left(\frac{\log\log N}{\log N}N+\frac{(1-\alpha)^2(1-\delta)^2}{ \epsilon^2}\right)$ as proved in Theorem~\ref{thm:perf_cpt}, our algorithm reduces the time overhead by a factor of up to $\log N$ over the best algorithm in the literature, advancing the state of the art on missing tag identification. Compared to the theoretical performance limit we derive, there is still room for improvement over our algorithm. We hope that our work will stimulate further algorithmic study that can narrow or even fill the performance gap.

\textbf{Roadmap}. The rest of the paper is articulated as follows. Section~\ref{II} introduces the system model and formulates the problem. Theoretical performance limit is derived in Section~\ref{III}. Section~\ref{IV} quantifies major existing RFID missing tag identification algorithms. We present our proposition, termed as collision-partition tree-based (CPT) algorithm, in Section~\ref{V}. Extensive experiments are conducted in Section~\ref{VI}. Section~\ref{sec:related_work} summarizes related work and analyzes their limitations. Section~\ref{IX} concludes the paper. 


\section{System Model and Problem Formulation}
\label{II}

Consider a time-synchronized RFID system composed of a reader and a set $\cal N$ of $N$ tags, each characterized by a unique $M$-bit ID ($M=96$ in the current standard). The reader communicates with tags via the standard  Framed Slotted Aloha (FSA) protocol. The time slots are divided into \emph{empty}, \emph{singleton} and \emph{collision} slots: in empty slots, no tag responds; in singleton slots, only one tag responds; a slot is called a $k$-collision slot if $k$ tags responds in this slot; collision slots and singleton slots are collectively referred to as busy slots. Slots differ in their length depending on the number of bits transmitted within the slots. Typically, we distinguish three slot lengths: $t_{s}$ denotes the length of a short slot containing a single bit, $t_{t}$ denotes the length of a tag slot containing a tag ID (i.e.\ $M$ bits), $t_{l}$ denotes the length of a long slot containing more information transmitted between the reader and tags.

\begin{table}[t]
	\centering  
	\caption{Main notations}  
	\label{table1} 
	\begin{tabular}{ll}  
		\toprule  
		$\mathcal{N}$ & Tag set \\ [1pt]
		$N$  & Cardinality of $\mathcal{N}$, $N=\mid \mathcal{N}\mid$ \\ [1pt]
		$\mathbb{A}$ & Set of missing tags  \\ [1pt]
		$\mathbb{B}$ & Set of tags returned by the algorithm \\ [1pt]
		$\alpha$ & $\mid \mathbb{A}\mid/N$ \\ [1pt]
		$\epsilon,\delta$ & Identification accuracy requirement \\ [1pt]
		$M$ & Tag ID length  \\ [1pt]
		$H(\cdot)$ & Hashing function \\ [1pt]
		$s$ & Hashing seed \\ [1pt]
		$I_t$ & ID of tag $t$ \\  [1pt]
		$I_t'$ & Pseudo-ID of tag $t$ \\  [1pt]
		$I_t'(b)$ & $b$-th bit in $I_t'$ \\  [1pt]
		$h$ & Height of a tree \\  [1pt]
		\bottomrule
		\vspace{-0.7cm}
	\end{tabular}
\end{table}

\begin{figure}[htbp]
  \centering
  \includegraphics[width=3.4in]{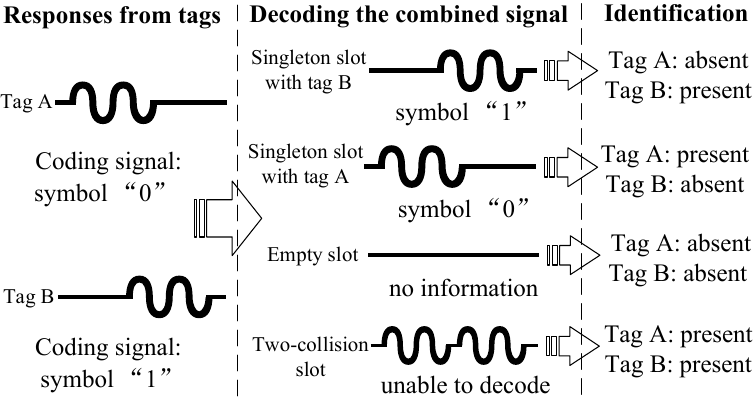}
  \caption{Illustration of Manchester coding.}
  \label{fig:manchester}
  \vspace{-0.2cm}
\end{figure}

At the physical layer, we adopt the standard Manchester coding widely applied in RFID systems~\cite{Roux_2002}. As depicted in Fig.~\ref{fig:manchester}, the main motivation of using Manchester coding~\cite{Roux_2002} is its capability of checking the tag presence from a $2$-collision slot when the two tags involved in the collision transmit different bits, i.e., ``1'' and ``0'', respectively. More specifically, ``0'' and ``1'' are encoded by a positive and negative signal transitions, respectively. If 2 tags simultaneously reply ``0'' and ``1'' to the reader, then the positive and negative transitions interfere with each other, which generates 4 distinct combined signals, from which we can identify the missing tag. 


\begin{figure}[htbp]
  \centering
  \includegraphics[width=3.45in]{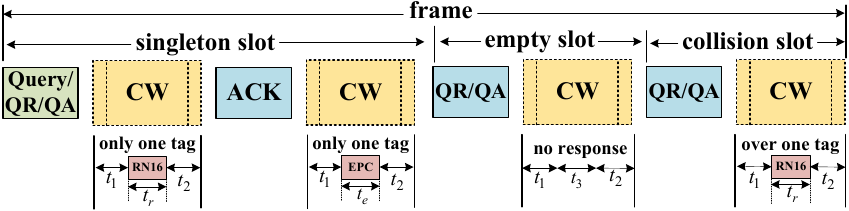}
  \caption{Illustration of the reader-tag dialogue. \emph{Query}, \emph{QueryRep} and \emph{QueryAdj} are used to initialize a new frame, among which \emph{QueryRep} and \emph{QueryAdj} are broadcast by the reader in tag identification. \emph{ACK}, \emph{CW}, \emph{RN16} and \emph{EPC} are the acknowledge command, the continuous ware, the 16-bit random number and the electronic product code, respectively.}
  \label{fig:command}
  \vspace{-0.2cm}
\end{figure}

We next briefly describe the tag identification process in RFID systems. As illustrated in Fig.~\ref{fig:command}, the tag identification process consists of multiple frames, each of which includes three categories of slots. According to the EPC C1 Gen2 specification~\cite{JSu_2020}, the reader initializes a frame or a slot by broadcasting \emph{Query} or \emph{QR} commands, respectively. With the received \emph{Query} command, each tag generates a 16-bit random number \emph{RN16}, and uses it as the hashing seed to generate a time counter \emph{TC}. When \emph{TC=0}, each tag transmits its \emph{RN16} to the reader. Upon receiving a legal \emph{RN16} from a tag, the reader acknowledges the \emph{RN16} by returning an \emph{ACK} command to this tag, and waits for its response \emph{EPC}. The reader then checks the slot by decoding the received \emph{EPC}. In Fig.~\ref{fig:command}, $t_{1}$ is the duration between reader transmission and tag reply, $t_{2}$ is the time needed for decoding the tag response, $t_{3}$ is the time the readers waits for for the tag response, $t_{r}$ and $t_{e}$ are the transmission time for \emph{RN16} and \emph{EPC}, respectively.

We conclude this section by formulating the missing tag identification problem.

\begin{problem}[Missing tag identification]
Let $\mathbb{A}$ denote the set of missing tags. Given the system parameters $\epsilon$ and $\delta$, we seek a missing tag identification algorithm to minimize the expected execution time while satisfying the following identification accuracy requirement: 
\begin{equation*}
P\left\{\frac{\mid \mathbb{A} \cap \mathbb{B}\mid}{\mid \mathbb{A}\mid}\ge1- \epsilon \right\} \ge 1- \delta,
\end{equation*}
where $\mathbb{B}$ denotes the tags returned by the algorithm.
\end{problem}




\section {Algorithm-independent Performance Bound}
\label{III}

In this section, we establish the lower-bound of the average execution time of any missing tag identification algorithm. To this end, we derive the lower-bound as functions of $\alpha$, $\delta$, $\epsilon$, and $N$ in Lemma~\ref{lem:1} and Lemma~\ref{lem:2}, respectively. We then merge them to establish the global lower-bound in Theorem~\ref{thm:bound}.

\subsection{Lower-bound as a function of $\alpha$, $\delta$ and $\epsilon$}

We derive the first lower-bound by leveraging Cook-reduction~\cite{Cook_1971}. We say that a problem $\cal A$ can be Cook-reduced to another problem $\mathcal{B}$ iff there exists a polynomial-time oracle machine $\mathcal{M}$ that can invoke every function $f$ that can solve $\mathcal{A}$ as a subroutine to solve $\mathcal{B}$. This implies that $\mathcal{B}$ is harder than $\mathcal{A}$ and a lower-bound on the overhead of any solution to $\mathcal{B}$ could be derived from that to $\mathcal{A}$. In our reduction, given the missing tag identification problem as $\mathcal{B}$, we select the \emph{Hamming Distance Estimation} (HDE)~\cite{Chakrabarti_2010} problem as $\mathcal{A}$. The HDE problem is a $2$-party communication complexity problem, where Alice and Bob receive $m$-bit binary strings $x \in \{0,1\}^{m}$ and $y \in \{0,1\}^{m}$ as input, respectively. They are required to count the number of the distinct bits between $x$ and $y$, with a given $(\epsilon,\delta)$ estimation accuracy requirement, while minimizing the number of the bits exchanged between them. A theoretical result from~\cite{ChenB_2013} shows that for $\epsilon<2/\sqrt{m}$ and $\delta=1/3$, solving the HDE problem for any protocol needs to exchange at least $\Omega(m)$ bits between the two parties.

Technically, given any missing tag identification algorithm $\mathcal{P}$, we use $\mathcal{P}$ as a building block to design another algorithm $\mathcal{P'}$ to solve the HDE problem. To this end, we need to establish a theoretical connection between the outputs of $\mathcal{P'}$ and $\mathcal{P}$, such that $\mathcal{B}$ is solved by $\mathcal{P'}$ once $\mathcal{A}$ is solved by $\mathcal{P}$.  Specifically, we make the number of missing tags equal the Hamming distance between $x$ and $y$. Alice and Bob can be seen as two players at the reader and tag sides, respectively. They virtually execute $\mathcal{P}$ on $N$ tags one by one, so as to obtain the two strings $x \in \{0,1\}^{m}$ and $y \in \{0,1\}^{m}$, which are inputs of  $\mathcal{P'}$. Given an arbitrary tag $i$, with $x[i]$ and $y[i]$ as its corresponding inputs for $\mathcal{P'}$, it is considered \emph{present} and if $x[i]=y[i]$, and \emph{missing} otherwise. By this way, we can design an algorithm $\mathcal{P'}$ solving $\mathcal{B}$ by invoking $\mathcal{P}$. Hence, Alice and Bob can solve the HDE problem as long as all missing tags are identified.

To successfully estimate the Hamming distance, Alice and Bob need to guarantee the correctness of the two strings $x \in \{0,1\}^{m}$ and $y \in \{0,1\}^{m}$. Since Alice operates at the reader side, she does not know which slots are empty before the simulated execution of $\mathcal{P}$, while Bob at the tag side can easily acquire that. This requires Bob to point out all empty slots, and further marks each tag (i.e.\ a missing tag) assigned to these empty slots with a bit different from that given by Alice. With the received response string $y \in \{0,1\}^{m}$ from Bob, Alice compares it with her string $x \in \{0,1\}^{m}$ to get the final result. Next, we describe how to determine whether a slot is empty. Specifically, given an arbitrary slot $S$ to which a set of tags $T$ map, Bob first calculates a fingerprint string  $\{y[t],t\in T\}$ and transmits it to Alice. Alice also computes a fingerprint string as $\{x[t],t\in T\}$. On receiving $\{y[t],t\in T\}$, Alice compares it with $\{x[t],t\in T\}$ to check whether $s$ is empty: only if $x[t]\ne y[t], \forall t\in T$, all tags in $T$ are missing and $S$ is empty. By repeating the operation on each slot, Alice and Bob can identify all the empty slots and then output all missing tags. 

The above Cook-reduction leads to the following lemma. The core technique to prove it is to choose $m=(1-\delta)^2(1-\alpha)^2/\epsilon^2$ and establish the relationship between the estimation accuracy and the number of required bits to achieve it. Due to space limit, we provide the proof of two major lemmas/theorems in the Appendix while omitting the proof details of others, some of which follow similar analysis.

\begin{lemma}
\label{lem:1}
No algorithm can meet the accuracy $(\epsilon,\delta)$ within $o\left(\frac{(1-\delta)^2(1-\alpha)^2}{\epsilon^2 \log \frac{(1-\delta)(1-\alpha)}{\epsilon}}\right)$ time, $\forall \epsilon\ge (1-\alpha )(1-\delta)/\sqrt{N},\delta<1/3$.
\end{lemma}

\subsection{Lower-bound as a function of $N$}
We derive the second lower-bound by leveraging Yao's minimax principle~\cite{Yao_1977}. To prove a lower-bound on the expected execution time of any randomized algorithm $\mathcal{P}$ with $\epsilon \le 1/2$, we consider a deterministic algorithm $\mathcal{P}^\ast$. Yao's minimax principle claims that the performance bound of a randomized algorithm can be derived from that of a deterministic algorithm over the worst input distribution. 


Technically, we first consider an input distribution, where $\mid \mathbb{A}\mid=\alpha N$ missing tags are uniformly distributed among $N$ tags.\footnote{For conciseness, throughout this paper, we asymptotically treat large real numbers such as $\alpha N$ as integers. In practice, a flooring or ceiling operation needs to be performed depending on the particular context.} 
$\mathcal{P}^\ast$ takes at least $\log N$ bits to encode all the tags.
Since each bit can be used to check at most $\mathcal{O}(1)$ tags from a single slot, $\mathcal{P}^\ast$ with $\log N$ bits can check at most $\mathcal{O}(\log N)$ tags from a single slot.
Recall the input distribution and $\epsilon \le 1/2$, we have that any $\log N$ tags cannot be identified within the error rate $\epsilon$ from a single slot. Hence, $\mathcal{P}^\ast$ cannot achieve the error rate $\epsilon$ within $o(N/\log N)$ slots.
To analyze the impact of $\delta$ on the lower-bound, we further consider
$N$ probabilistic inputs. For input $i$, the number of tags is $i$ and the number of missing tags is $\alpha i$. Each input  appears with the same probability $1/N$. To ensure $\delta < 1/3$, $\mathcal{P}^\ast$ needs to achieve $\epsilon$ accuracy on at least $2/3$ of all the inputs. Hence, $\mathcal{P}^\ast$ needs to give at least $2N/3$ different outputs, which take $\Omega(\log N)$ bits to encode. If $\mathcal{P}^\ast$ consumes only $o(\log N)$ bits, it can generate at most $o(N)$ different outputs, which is obviously not enough. Combining the above analysis, we have that no algorithm ${\cal P}^\ast$ can achieve $(\epsilon,\delta)$ identification accuracy within $o(N/\log N+\log N)=o(N/\log N)$ time, as stated in Lemma~\ref{lem:2}.

\begin{lemma}
No algorithm can meet the accuracy $(\epsilon,\delta)$ within $o\left(N/\log N\right)$ time, $\forall\epsilon\le 1/2,\delta<1/3$.
\label{lem:2}
\end{lemma}

Lemma~\ref{lem:1} and~\ref{lem:2} readily leads to a performance lower-bound $\Theta \left(\frac{N}{\log N}+\frac{(1-\delta)^2(1-\alpha)^2}{\epsilon^2 \log \frac{(1-\delta)(1-\alpha)}{\epsilon}}\right)$, as stated in Theorem~\ref{thm:bound}.

 

\begin{theorem}[Algorithm-independent performance bound]
No missing tag identification algorithm can achieve $(\epsilon,\delta)$ identification quality within $o\left(\frac{N}{\log N}+\frac{(1-\delta)^2(1-\alpha)^2}{\epsilon^2 \log \frac{(1-\delta)(1-\alpha)}{\epsilon}}\right)$ time, $\forall \epsilon \in [\frac{(1-\alpha )(1-\delta)}{\sqrt{N}}, \frac{1}{2}]$, $\delta \in [0,\frac{1}{3})$.
\label{thm:bound}
\end{theorem}

\begin{remark}
To complete the result of Theorem~\ref{thm:pcmti}, we note that for $\epsilon<(1-\alpha )(1-\delta)/\sqrt{N}$, by applying the results derived in~\cite{Chakrabarti_2010}, we can prove that the trivial deterministic polling algorithm outperforms any randomized algorithm. Hence, the performance bound in this case sums up to $\Theta(N)$. Therefore, our subsequent analysis focuses on the case $\epsilon\ge(1-\alpha )(1-\delta)/\sqrt{N}$. 
\end{remark}

\section{Quantifying Major Existing Missing Tag Identification Algorithms}
\label{IV}

Given the performance limit derived in Theorem~\ref{thm:bound}, in this section we quantify the major existing missing tag identification algorithms. Due to space limit, we present our analysis for a representative algorithm PCMTI~\cite{ZhangL_2017} stated in Theorem~\ref{thm:pcmti}, and summarize the results for others in Tab.~\ref{tab:comparison}.

\begin{table}[t]
	\centering
	\caption{Comparison of major missing tag identification algorithms in the literature} 
	\label{tab:comparison}
	\begin{tabular}{ll}  
        \toprule    
		Algorithm & Expected execution time \\ [1pt] 
		\midrule 
		IIP, THP~\cite{LiT_2010,LiT_2013} & $\Theta \left(N +\frac{(1-\alpha)^2(1-\delta)^2}{ \epsilon^2}\right)$ \\ [1pt]
		P-MTI~\cite{ZhengY_2013} & $\Theta \left(N \log N+\frac{(1-\alpha)^2(1-\delta)^2}{ \epsilon^2} \right)$ \\ [1pt]
		MMTI~\cite{LiuX_2014} & $\Theta \left(N \log N+\frac{(1-\alpha)^2(1-\delta)^2}{ \epsilon^2} \right)$ \\[1pt]
		SFMTI~\cite{LiuX_2015} & $\Theta \left(N+\frac{(1-\alpha)^2(1-\delta)^2}{ \epsilon^2}\right)$ \\[1pt]
		ProTaR~\cite{ShaoC_2015} & $\Theta \left(N+\frac{(1-\alpha)^2(1-\delta)^2}{ \epsilon^2}\right)$ \\[1pt]
		PCMTI~\cite{ZhangL_2017} &  $\Theta \left(N+\frac{(1-\alpha)^2(1-\delta)^2}{ \epsilon^2}\right)$ \\[1pt]
		CPT (our solution)  &  $\Theta \left(\frac{\log\log N}{\log N}N+\frac{(1-\alpha)^2(1-\delta)^2}{ \epsilon^2}\right)$ \\
		Theoretical performance limit & $\Theta\left(\frac{N}{\log N}+\frac{(1-\delta)^2(1-\alpha)^2}{\epsilon^2 \log \frac{(1-\delta)(1-\alpha)}{\epsilon}}\right)$ \\
		\bottomrule  
	\end{tabular}
	\vspace{-0.4cm}
\end{table}

\begin{theorem}
PCMTI needs $\Theta (N+(1-\alpha)^2(1-\delta)^2/\epsilon^2)$ time to achieve $(\epsilon,\delta)$ identification quality, $\forall \epsilon \in [(1-\alpha )(1-\delta)/\sqrt{N}, 1/2]$, $\delta \in [0,1/3)$. 
\label{thm:pcmti}
\end{theorem}

PCMTI features two breakthrough techniques: the pair-reply and $2$-collision resolving strategies. The former can be used to check up to two tags in one short-response slot simultaneously, while the latter upgrades the utilization efficiency of each frame by rearranging two consecutive singleton slots as a $2$-collision slot. According to such design of PCMTI, we give the proof sketch below.
Following our technique in the analysis of Theorem~\ref{thm:bound}, we derive the performance as a function of $\epsilon$, $\delta$, $\alpha$, and $N$, respectively. 

\textbf{Performance bound as a function of $\epsilon$, $\delta$ and $\alpha$.} We construct a stochastic process based on  observations from each slot. Specifically, we treat each slot as a single observation, and denote $X$ as a random variable representing an observation from an arbitrary slot. $X=1$ if there are missing tags identified in the slot and $0$ otherwise. Obviously, $X$ follows Bernoulli distribution. We further denote $\overline{X}=\frac{1}{n}\sum\nolimits_{i=1}^{n}X_{i}$ as another random variable representing the mean of $n$ independent observations, where $X_{i}$ denotes the $i$-th observation of $X$. Following the law of large numbers~\cite{Etemadi_1981}, we obtain  $\mathbb{E}[\overline{X}]=\mathbb{E}[X]$ and $\sigma (\overline{X})=\sigma(X)/\sqrt{n}$. We further define $Z=(\overline{X}-\mu)/\sigma$, where $\mu=\overline{X}$ and $\sigma (Z)=\sigma (\overline{X})$. It follows from the central limit theorem~\cite{Hoeffding_1985} that asymptotically $Z$ follows the standard normal distribution. We compute the one-sided confidence interval $C$ satisfying ${P}\{Z\ge C=\overline{X}-\frac{\sigma(Z)}{\sqrt{r}}t_{\delta}(r-1)\}\ge 1-\delta$, where $t_{\delta}(r-1)$ denotes a $t$-distribution parameter, $r$ represents the total number of trials for $Z$, and $\frac{\overline{X}-\mu}{\sigma(Z)/\sqrt{r}} \sim t_{\delta}(r-1)$. After some algebraic operations, we obtain $n=\Theta\left(\frac{(1-\alpha)^2(1-\delta)^2}{\epsilon^2}\right)$, which corresponds to the expected time overhead to achieve the required identification accuracy. The detailed demonstration of this part is given in Appendix B.


\textbf{Performance bound as a function of $N$.} The execution time  is dominated by the the number of slots containing information multiplied by the quantity of information transmitted in each of them. Major existing solutions exploit a slot to check only one tag, thus requiring $\Theta(N)$ slots in total, even with an error rate $\epsilon\ge(1-\alpha )(1-\delta)/\sqrt{N}$. To break the bottleneck, PCMTI employs a pair-reply strategy that can simultaneously check up to $2$ tags in each slot with $1$-bit response. 
By only retaining $2$-
collision slots, PCMTI guarantees that there must be missing
tags in an empty or singleton slot.
To analyze its performance bound, we focus on the probability $P_{t_{m}}$ that represents $t_{m}$ missing tags have been verified before the first $n_{f}$ slots. 
Let ${P}_{01}$, ${P}_{10}$, and ${P}_{11}$
denote the probabilities that a missing tag is assigned to a slot, a missing tag and a present tag are assigned to a slot, and two missing tags are assigned to a slot, respectively. To this end, the probability that an arbitrary missing tag is identified in a slot can be formalized by 
${P}_{m}={P}_{01}+{P}_{10}+{P}_{11}$. Naturally, the probability that this tag has not been verified before the first $n_{f}$ slots is ${P}_{t}=(1-{P}_{m})^{n_{f}}$, thus $P_{t_{m}}=\binom{\alpha N}{t_{m}} (1-{P}_{S})^{t_{m}} {P}_{S}^{(\alpha N-t_{m})}$. We terminate the protocol when it guarantees $P\{t_{m} \ge (1-\epsilon) \mid \mathbb{A}\mid \}\ge 1- \delta$, from which  we obtain $n_{f}=\Theta(N/\log N)$. Besides, to rearrange two reserved singleton slots as a $2$-collision slot at full steam, PCMTI performs an information broadcasting operation within a complexity space containing $\Theta({N})$ bits for each slot, which consumes $\Theta({\log N})$ bits. We thus obtain an overall time overhead of $\Theta(N/\log N)\cdot\Theta({\log N})=\Theta(N)$.

\section{Missing Tag Identification Based on \underline{C}ollision-\underline{P}artition \underline{T}ree (CPT)}
\label{V}

Armed with the theoretical performance limit derived in Sec.~\ref{III}, we design a novel missing tag identification algorithm  approaching the theoretical limit and outperforming all the solutions in the literature, as illustrated in Tab.~\ref{tab:comparison}.  

The key innovation in our algorithm is a novel data structure we develop, termed as collision-partition tree (CPT). Different from the conventional query tree commonly used in the literature constructed based on tag ID prefix, our CPT is built upon a subset of bits in the pseudo-ID of tags. The pseudo-IDs can be regarded as a condensed hashprint of tag IDs with the property that any pair of tags in the system have distinct pseudo-IDs. Relying on a selected subset of bits in tag pseudo-IDs leads to shorter and more balanced tree structure (cf. Fig.~\ref{UI}) and hence reduces the time complexity in parsing the entire tree. 
Our second improvement over the state-of-the-art solutions is to employ Manchester coding such that a single slot can convey the information enabling us to check the presence of two tags. 

Our algorithm, termed as CPT, is composed of three steps. Step 1 generates the pseudo-IDs for all the tags in the system. Step 2 establishes the CPT based on tag pseudo-IDs instead of tag ID prefix. Step 3 employs the Manchester encoding to identify the missing tags from the leaves of the CPT constructed in Step 2. 

\subsection{Our Algorithm}

\begin{figure}
  \begin{center}
  \includegraphics[width=3.5in]{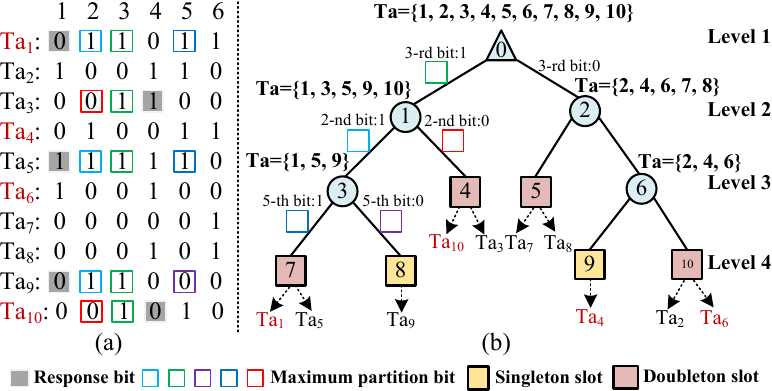}
  \caption{(a) Pseudo-IDs of tags; (b)  the constructed CPT}\label{UI}
  \end{center}
  \vspace{-0.5cm}
\end{figure}

\textbf{Step 1: constructing tag pseudo-IDs.} As the height of the CPT grows with the length of tag IDs, in Step 1, the reader computes a pseudo-ID for each tag based on its tag ID such that (1) the pseudo-ID for each tag in the system is unique, (2) the length of pseudo-IDs is much shorter than the original tag ID. To this end, for each tag $t$ with ID $I_t$, we set its pseudo-ID $I_t'$ as the last $2\log N$ bits of $H(I_t,s)$ where $s$ is a random seed and $H()$ is a uniform hashing function. We then denote $I_t'(b)$ as the $b$-th bit in $I_t'$. As we do not know the exact value of $N$, we can use an estimation of its upper-bound to set the length of the pseudo-IDs. We repeat the iteration until the pseudo-IDs of any pair of tags are distinct. The following lemma formally proves that we can successfully construct a pseudo-ID for each tag w.h.p.\footnote{By w.h.p., we mean with probability $1-N^{-a}$ for any $a>0$.} The proof follows from the famous birthday paradox. Fig.~\ref{UI}(a) illustrates the pseudo-IDs generated for $N=10$ tags.

\begin{lemma}
We can successfully construct a distinct pseudo-ID for each tag in $\mathcal{O}(1)$ iterations w.h.p.
\end{lemma}

\textbf{Step 2: constructing CPT.} A CPT is composed of a set of leaves and internal nodes. 
\begin{itemize}
    \item \textbf{Leaves.} Each tag maps to a leaf. Each leaf has up to $2$ tags mapping to it.
    \item \textbf{Internal nodes.} Each internal node maps to a bit position in the pseudo-IDs, allowing us to divide the tags into $2$ subsets. Take Fig.~\ref{UI} as an example. The root (node $0$) maps to the $3$rd bit in the tag pseudo-IDs. Hence, the tags $1,3,5,9,10$ map to the left sub-tree of node $0$, while the tags $2,4,6,7,8$ map to the right sub-tree.
\end{itemize}

The CPT is constructed from the root to the leaves. At the starting point, all the tags are associated to the root. The reader selects the bit in the pseudo-ID that can divide the tags in the most balanced way as described above, breaking ties randomly. It then repeats this process iteratively until reaching the leaves.



\textbf{Step 3: identifying missing tags.} Let $h$ denote the height of the CPT constructed by the reader. $h_{min}$ and $h_{max}$ denote the maximal and minimal heights of the leaves in the constructed CPT. The reader iterates from $h_{min}$ to $h_{max}$. In each iteration, it checks the presence of each leaf of height $h$, until all the leaves are parsed. 
\begin{itemize}
    \item To parse the first leaf $i$, the reader broadcasts the path from the root to $i$. For example, to parse the leaf $7$ in Fig.~\ref{UI}, the reader broadcasts the path $\{(3,1), (2,1), (5,1)\}$ to reach it. Besides the path, the reader also broadcasts the position of the bit in the pseudo-ID that can distinguish the tags mapping to the parsed leaf. This is possible because (1) there are at most $2$ tags mapping to each leaf, (2) the reader can check the presence of each tag of the leaf by exploiting Manchester coding.
    \item To parse subsequent leaves, the reader does not need to broadcast the entire path. Instead, it can broadcast the path from the previously parsed leaf to the currently parsed leaf, thus reducing the quantity of broadcast information. For example, to parse the leaf $8$ after parsing its sibling leaf $7$, the reader broadcasts the path between them via node $3$ without restarting from the root.  
\end{itemize}

\begin{figure}
  \begin{center}
  \includegraphics[width=3.2in]{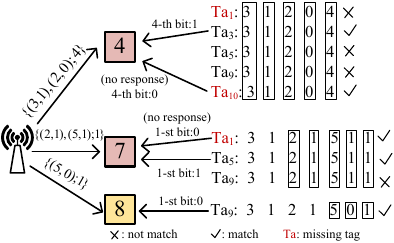}
  \caption{Illustration of Step 2: parsing the sub-tree rooted in node $1$ in the CPT in Fig.~\ref{UI}}\label{reader}
  \end{center}
\vspace{-0.9cm}
\end{figure}

\begin{example}
\label{ex:iden}
Fig.~\ref{reader} illustrates Step 3 for parsing the sub-tree rooted in node $1$ in the CPT in Fig.~\ref{UI}. The reader starts by parsing the highest leaf, i.e., node $4$. To this end, it broadcasts $\{(3,1), (2,0), 4\}$, requiring any tag $i$ whose pseudo-ID $I_i'$ satisfying $I_i'(3)=1$ and $I_i'(2)=0$ to reply by sending $I_i'(4)$ in this slot. Only $Ta_{3}$ and $Ta_{10}$ match the condition, so they response to the reader with $1$ and $0$, respectively. With Manchester coding, the reader can check that $Ta_{10}$ is missing. After that, $Ta_{3}$ and $Ta_{10}$ keep silent until the algorithm terminates. On the other hand, as $I_i(3)=0$ for $i=2,4,6,7,8$, the tags $Ta_{2}$, $Ta_{4}$, $Ta_{6}$, $Ta_{7}$ and $Ta_{8}$ also keep silent in the current phase. The reader then moves to parse node $7$. With $I_i'(2)=1$, the tags $Ta_{1}$, $Ta_{5}$, $Ta_{9}$ know that they are candidates for node $7$. On receiving $\{(5, 1); 1\}$, only $Ta_{1}$ and $Ta_{5}$ match the condition, thus they respond to the reader with $0$ and $1$, respectively. $Ta_{1}$ is identified as missing from this slot in the similar way. Finally, the reader can check the presence of $Ta_{9}$ by broadcasting $\{1\}$. To further improve efficiency at the execution level, the reader can aggregate multiple commands within a broadcast packet, as described in~\cite{Liu_2019}. 
\end{example}

\subsection{Theoretical Performance Analysis}

To analyze the performance of our algorithm, we take a similar analysis method used in Sec.~\ref{III} and~\ref{IV}, i.e., deriving the time overhead as functions of $N$, and  $\epsilon$, $\delta$ $\alpha$, respectively. As explained previously, we focus on the non-trivial setting where $\epsilon \in [(1-\alpha )(1-\delta)/\sqrt{N}, 1/2]$, $\delta \in [0,1/3)$. In what follows, we give the analysis sketch followed by the core theorem on the performance. Due to space limit, we omit the details of analysis.

\textbf{Performance bound as a function of $N$.} We can intuitively observe that $\mathbb{E}[h]$ is upper-bounded by $\log N$ following the fact that tag pseudo-IDs are hashprints and thus uniformly distributed.
We then construct an auxiliary graph for the CPT, denoted as $G \triangleq (V, E)$, where the set of vertexes $V$ and the set of edges $E$ are leaves and the shortest paths among leaves. Moving from one leaf to another in our algorithm is thus mapped traversing the corresponding edges. Clearly, $G$ contains $\Theta(N)$ edges, and the degree of each vertex is almost $2$. The reader queries one leaf to another by broadcasting the bit index in $2 \log N $-bit pseudo-IDs in binary form, thus requiring $\Theta(\log \log N)$ bits to encode these information along the edge between them.
This is equivalent to saying that each slot consumes $\Theta(\log \log N)$ bits for information broadcasting.
With $\epsilon\ge(1-\alpha )(1-\delta)/\sqrt{N}$ and $\delta <1/3$, $\mathbb{E}[h]$ is reduced by $\Theta(\log\log N)$, thus our algorithm requires  $\Theta(2^{\log N-\Theta(\log \log N)})=\Theta(N/\log N)$ slots on such case. In this context, our algorithm incurs roughly $\Theta (\log\log N) \cdot \Theta (N/\log N)=\Theta (N\log\log N/\log N)$ time complexity. 

\begin{figure*}[t]
\centering
\captionsetup{font={footnotesize}}
\subfigure[$\delta=0.1$, $\alpha=0.01$, $N=50000$]{
\begin{minipage}[t]{0.23\textwidth}
\includegraphics[width=1.75in]{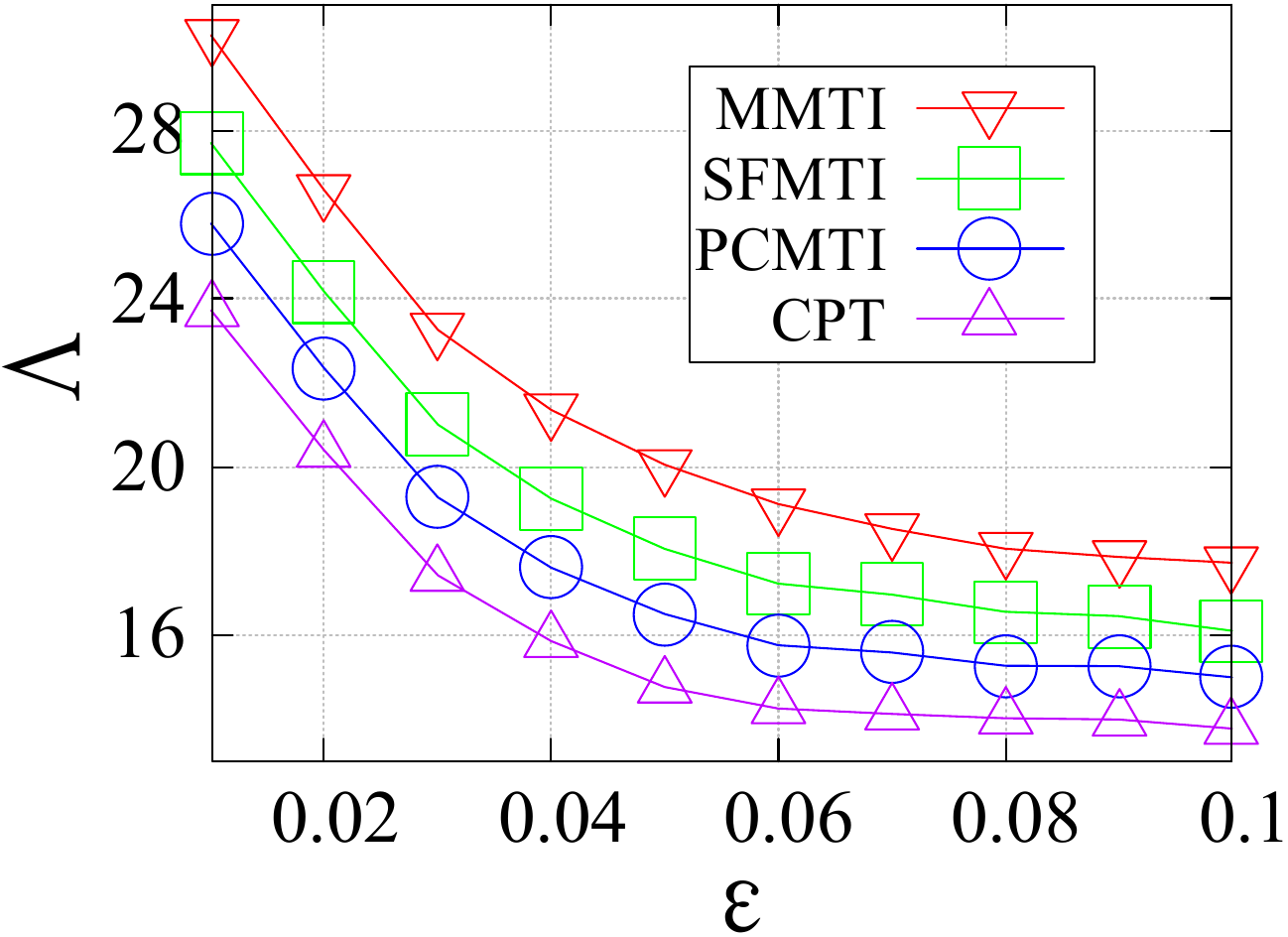}
\end{minipage}}
\subfigure[$\epsilon=0.01$, $\alpha=0.01$, $N=50000$]{
\begin{minipage}[t]{0.23\textwidth}
\includegraphics[width=1.75in]{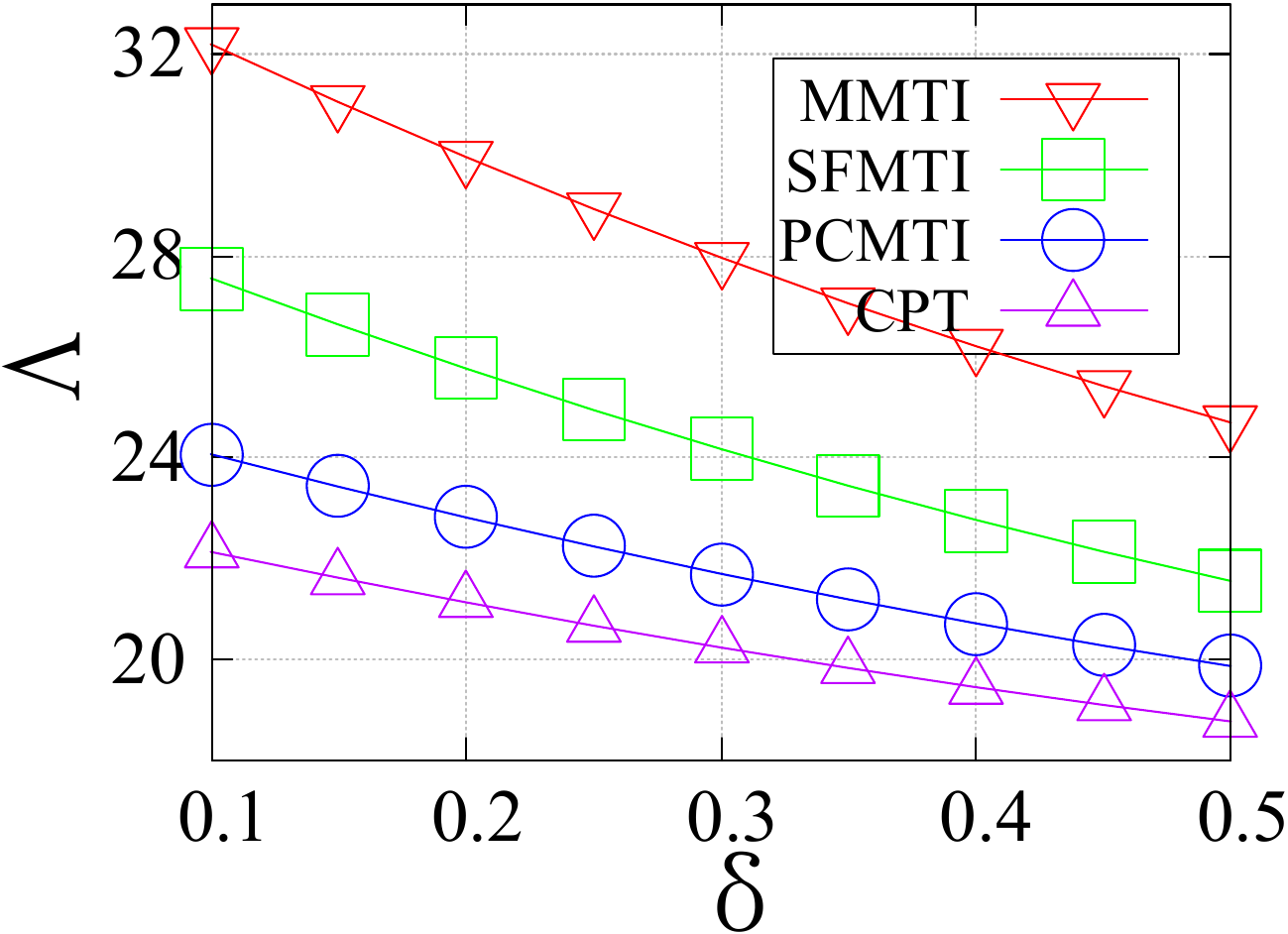}
\end{minipage}}
\subfigure[$\epsilon=0.01$, $\delta=0.1$, $N=50000$]{
\begin{minipage}[t]{0.23\textwidth}
\includegraphics[width=1.75in]{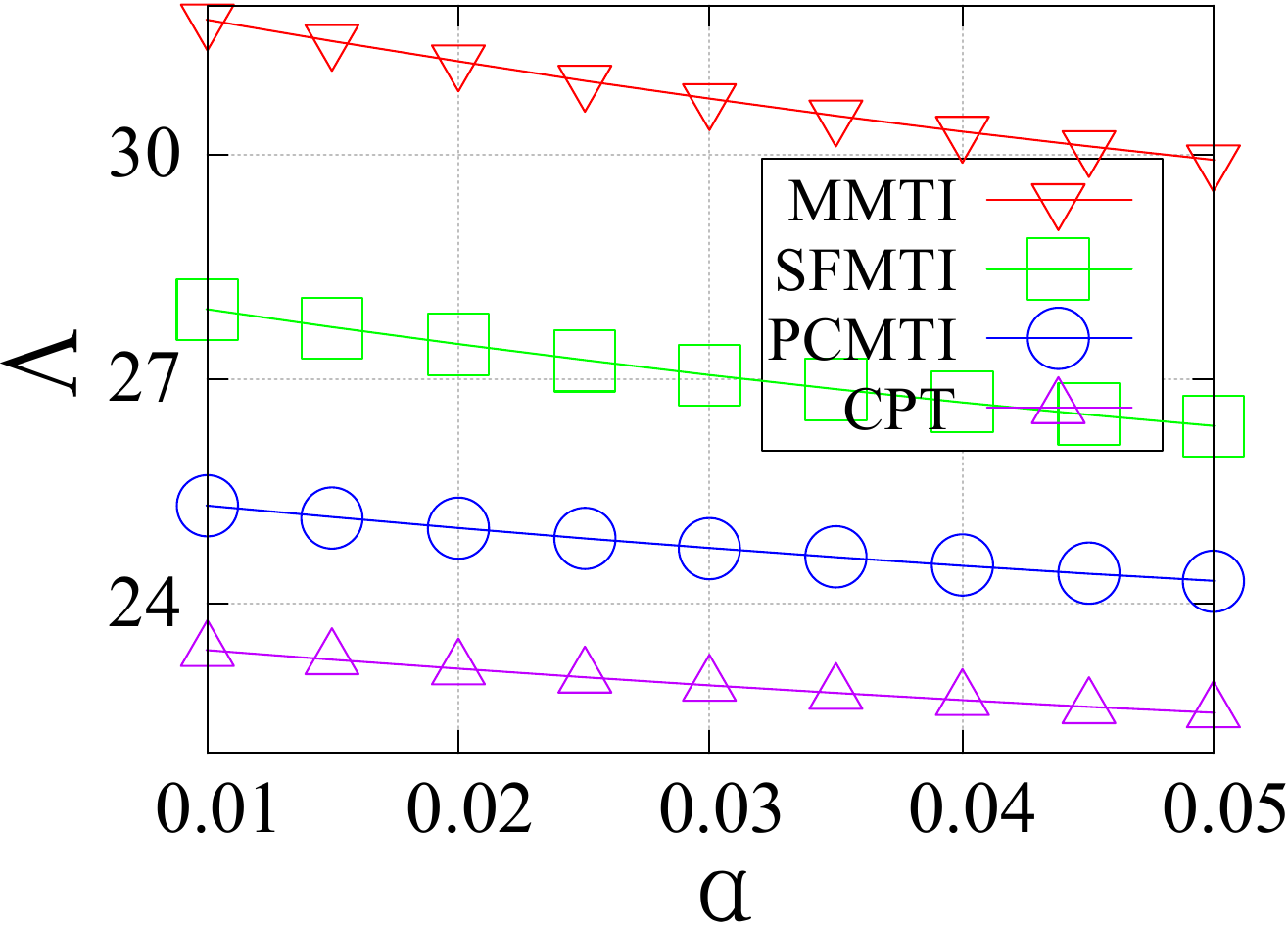}
\end{minipage}}
\subfigure[$\epsilon=0.01$, $\delta=0.1$, $\alpha=0.01$]{
\begin{minipage}[t]{0.23\textwidth}
\includegraphics[width=1.75in]{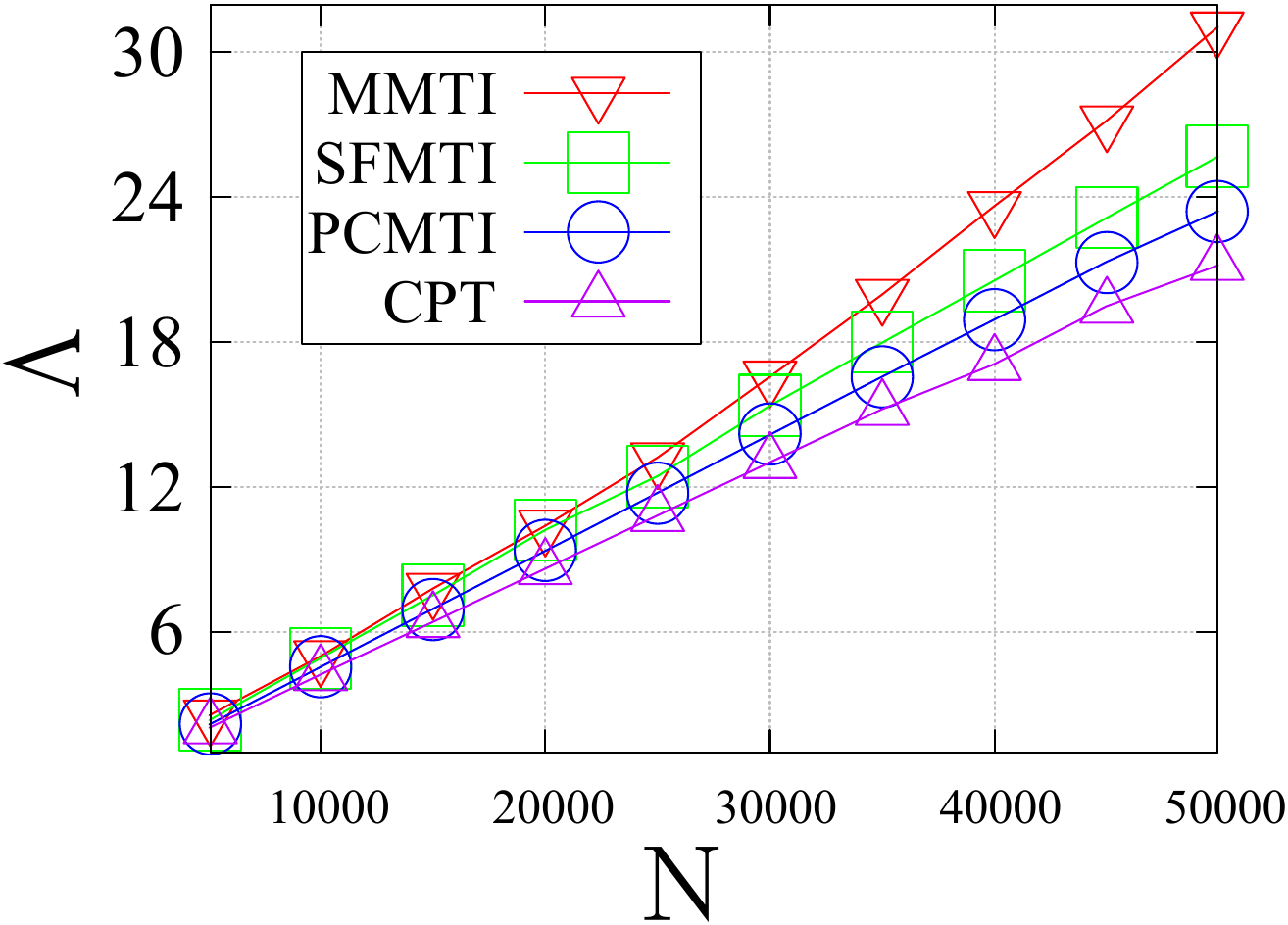}
\end{minipage}}
\caption{Performance comparison with different parameter settings from the first group of experiments.} \label{5}
\end{figure*}

\begin{figure*}[t]
\centering
\captionsetup{font={footnotesize}}
\subfigure[$\delta=0.1$, $\alpha=0.01$, $N=20000$]{
\begin{minipage}[t]{0.23\textwidth}
\includegraphics[width=1.75in]{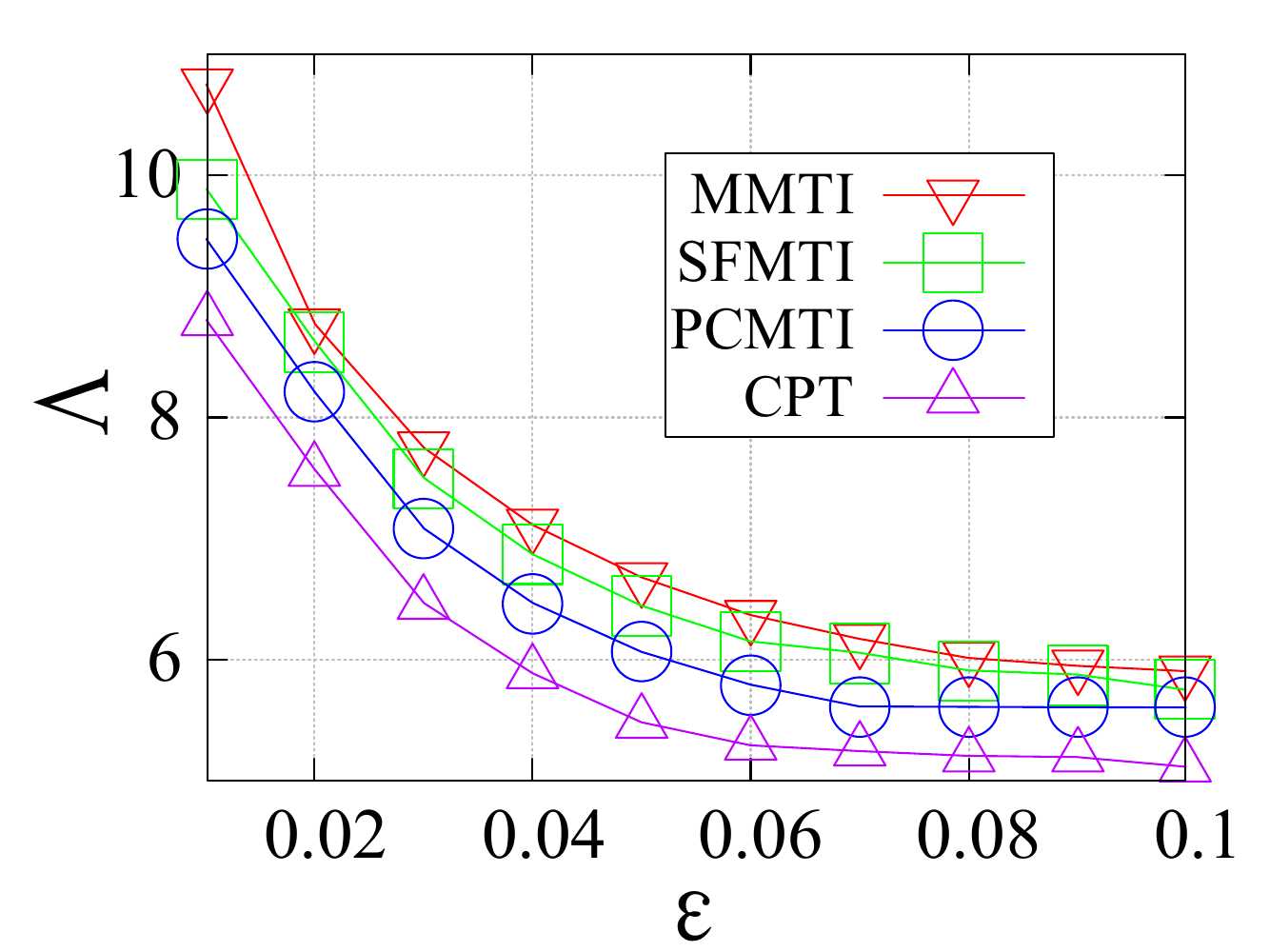}
\end{minipage}}
\subfigure[$\epsilon=0.01$, $\alpha=0.01$, $N=20000$]{
\begin{minipage}[t]{0.23\textwidth}
\includegraphics[width=1.75in]{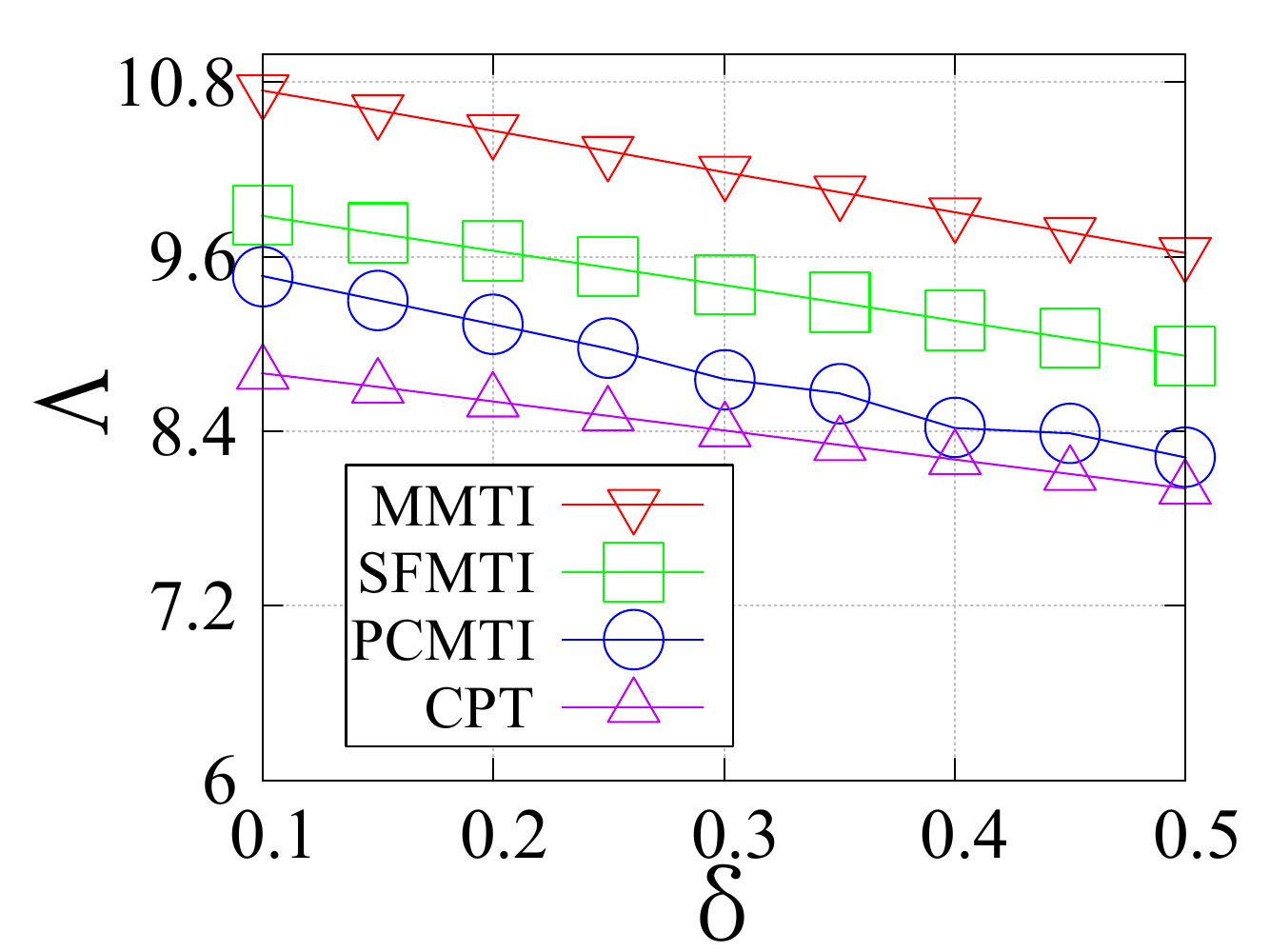}
\end{minipage}}
\subfigure[$\epsilon=0.01$, $\delta=0.1$, $N=20000$]{
\begin{minipage}[t]{0.23\textwidth}
\includegraphics[width=1.75in]{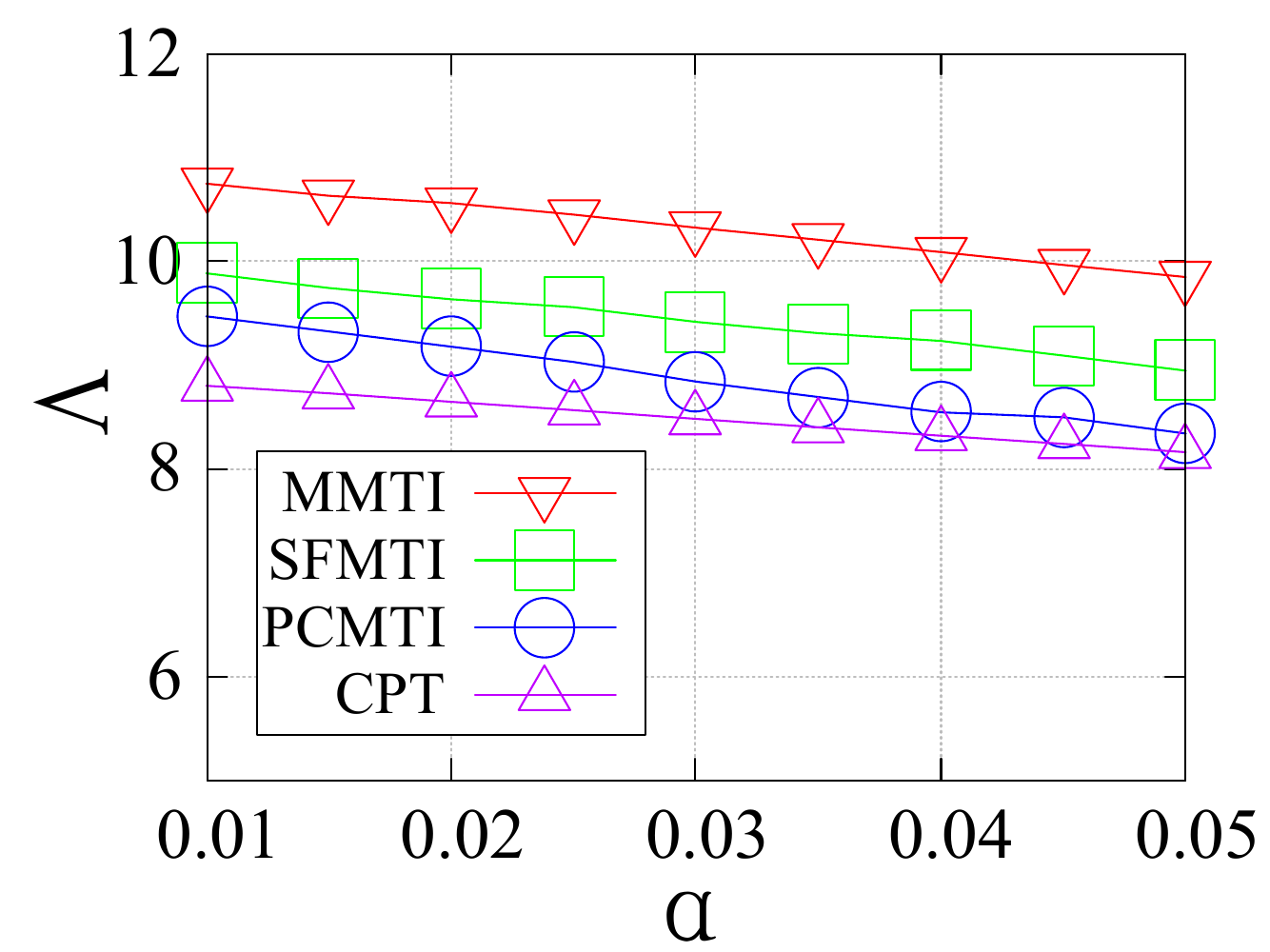}
\end{minipage}}
\subfigure[$\epsilon=0.1$, $\delta=0.1$, $\alpha=0.01$]{
\begin{minipage}[t]{0.23\textwidth}
\includegraphics[width=1.75in]{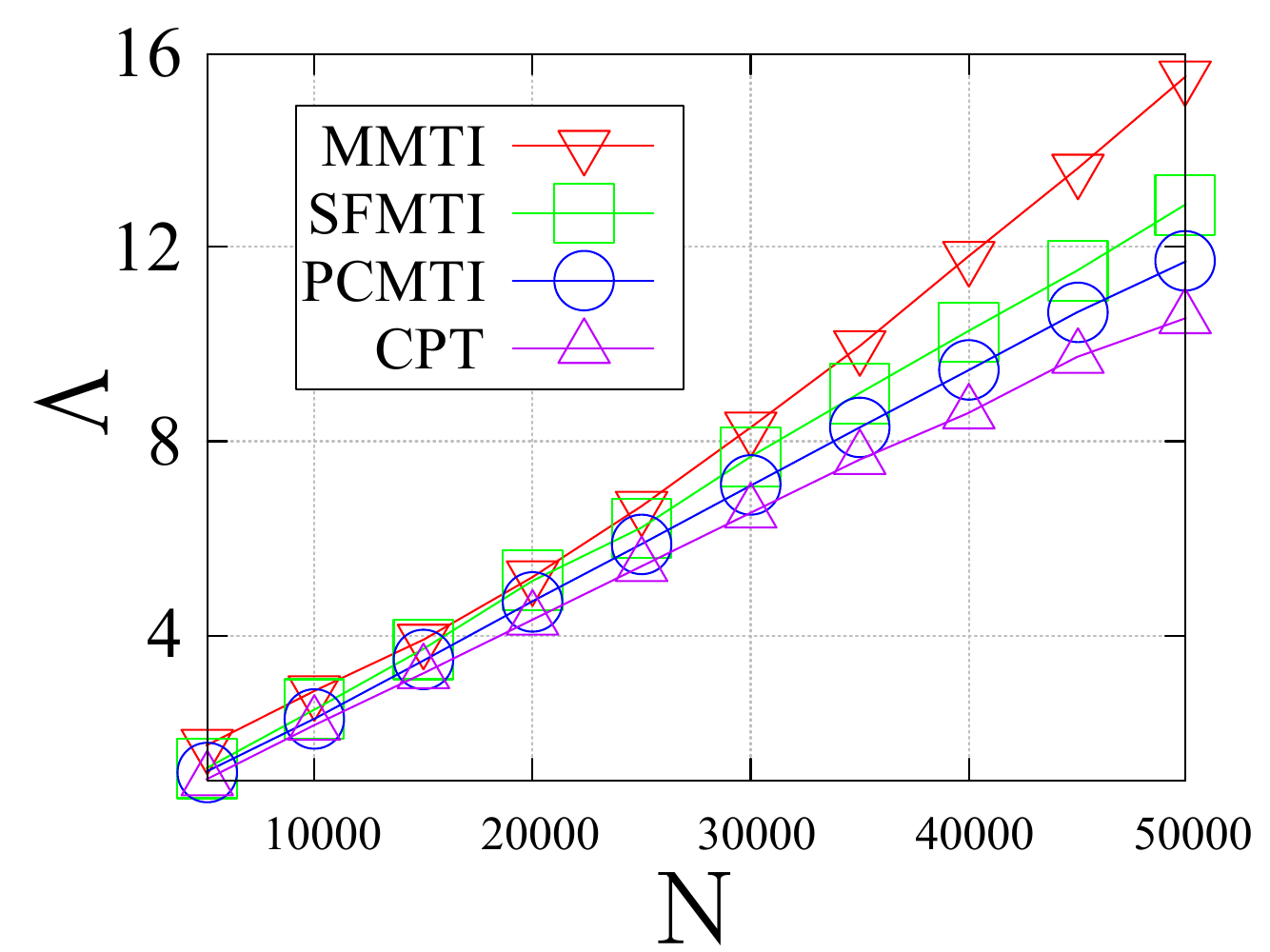}
\end{minipage}}
\caption{Performance comparison with different parameter settings from the second group of experiments.} \label{6}
\vspace{-0.3cm}
\end{figure*}

%

\textbf{Performance bound as a function of $\epsilon$, $\delta$ and $\alpha$.} By similar analysis as for PCMTI in Sec.~\ref{IV}, we can prove that our algorithm incurs $\Theta \left((1-\alpha)^2(1-\delta)^2/\epsilon^2\right)$ time overhead to guarantee $(\epsilon, \delta)$ accuracy. 


Combining the above analysis leads to the theorem below.




\begin{theorem} 
Our algorithm CPT takes $\Theta \left(\frac{N \log\log N}{\log N} +\frac{(1-\alpha)^2(1-\delta)^2}{\epsilon^2}\right)$ time to meet $(\epsilon,\delta)$ identification quality, $\forall \epsilon \in [(1-\alpha )(1-\delta)/\sqrt{N}, 1/2]$, $\delta \in [0,1/3)$. 
\label{thm:perf_cpt}
\end{theorem}

Theorem~\ref{thm:perf_cpt} demonstrates that our algorithm reduces the time overhead by a factor of up to $\log N$, more precisely, $\log N/\log \log N$, over the best algorithm in the literature, advancing the state of the art on missing tag identification. Nevertheless, compared to the theoretical performance limit we derive, there is still room for improvement over our algorithm. We hope that our work will stimulate further algorithmic study that can narrow or even fill the performance gap.

\section{Numerical Results}
\label{VI}
In this section, we conduct extensive simulations to further evaluate the numerical performance of our algorithm compared with three state-of-the-art algorithms: MMTI~\cite{LiuX_2014}, SFMTI~\cite{LiuX_2015}, and PCMTI~\cite{ZhangL_2017}. 

\subsection{Simulation Setup and Evaluation Metrics}

We configure the RFID system parameters as specified in Philips I-Code standard~\cite{jihong_2017_2}, where $t_{s}=0.4 ms$, $t_{t}=2.4 ms$ and $t_{l}=0.8 ms$. Therefore, the bit transmission rate from tags to the reader is set to $96/2.4=40 kbps$. The average execution time that each algorithm takes to achieve the identification accuracy $(\epsilon, \delta)$, denoted by $\Lambda$, is traced as the performance metric. For a fair comparison, we set up each algorithm using its optimal parameter configuration, and all our results are obtained from the average of $100$ simulation runs. We vary the parameters $\epsilon$, $\delta$, $\alpha$ and $N$ as illustrated in Fig.~\ref{5} and Fig.~\ref{6} to study their impact on $\Lambda$ for the simulated algorithms. 

\subsection{Simulation Results}

We perform two $2$ of experiments, each using a different set of parameter setting, as shown in Fig.~\ref{5} and Fig.~\ref{6}. Since we observe similar trends for both sets of experiments, our subsequent analysis mainly focuses on Fig.~\ref{5}  with sometimes a brief summary on Fig.~\ref{6} for comparison. Note that the unit of $\Lambda$ is seconds in both figures.
 
\textbf{Impact of $\epsilon$ on $\Lambda$.} From Fig.~\ref{5}(a), we observe that the execution time of all the simulated algorithms decreases more significantly under small $\epsilon$. This is consistent with our theoretical performance analysis. In comparison, our algorithm CPT outperforms the other solutions by $8-50\%$ over the best algorithms simulated in the experiment in terms of $\Lambda$. By comparing Fig.~\ref{5}(a) with Fig.~\ref{6}(a), we further observe that the performance improvement increases with $N$, which is also coherent with our theoretical finding and demonstrates the scalability of our algorithm.
 
\textbf{Impact of $\delta$ on $\Lambda$.} From  Fig.~\ref{5}(b), we observe that CPT significantly outperforms the other algorithms for all the values of $\delta$. The performance gap goes as large as $45\%$ w.r.t. MMTI. Moreover, CPT incurs less performance variation among the simulated algorithms, thus exhibiting the best stability among them. Similar to Fig.~\ref{5}(b), Fig.~\ref{6}(b) also illustrates large improvement on $\Lambda$ for CPT, with even less variation compared to Fig.~\ref{5}(b). 
 
\textbf{Impact of $\alpha$ on $\Lambda$.} 
 Similar to Fig.~\ref{5}(a), we observe from Fig.~\ref{5}(c) that the execution time of all the simulated algorithms decreases more significantly under small $\alpha$. This is also consistent with our theoretical performance analysis. CPT reduces $20-45\%$ execution time compared to the state-of-the-art solutions and demonstrates less performance variation and better scalability. The performance variation is more moderate with small $N$, as can be observed by comparing Fig.~\ref{5}(c) and Fig.~\ref{6}(c).
 
\textbf{Impact of $N$ on $\Lambda$.} As shown in Fig.~\ref{5}(d), with $N$ increasing from $5000$ to $50000$, the execution time of each algorithm increases significantly, demonstrating its high sensitivity over $N$. Nevertheless, we observe that the slope of CPT is smaller than the other algorithms. Particularly, we observe that the slope of CPT is decreasing in $N$, while the slope of other algorithms are either slightly increasing or at least non-decreasing for the entire range of the simulated values of $N$. As a result, we observe that with large $N$, the performance gain of CPT is more pronounced, demonstrating again the scalability of our algorithm.


Overall, we can draw the following conclusions from our numerical analysis: (1) $\epsilon$ and $N$ have a paramount influence on the execution time of each algorithm, while the impact of $\alpha$ and $\delta$ less important; (2) The execution time of each algorithm decreases more significantly under small $\epsilon$ while becoming less sensitive when $\epsilon$ approaches $0.1$; in contrast, the performance exhibits nearly constant degradation slope over $N$; (3) Our algorithm CPT outperforms all the simulated state-of-the-art solutions with non-negligible gain in terms of time overhead, scalability and stability. These conclusions are consistent with our theoretical analysis as shown in Tab.~\ref{tab:comparison}.

\section{Related Works}
\label{sec:related_work}

Missing tag identification has been regarded as one of the most fundamental tasks in RFID systems. There are two problem formulations related to missing tag identification. 

The first one is called missing event detection. We are given a threshold on the number of tags. A missing event occurs if the number of missing tags exceeds the threshold. The problem is to design an algorithm that can correctly report a missing event with a probability specified by the system. A handful of algorithms~\cite{ChiuC_2008, jihong_2017, LuoW_2011, LuoW_2012} have been proposed to address this problem without exactly determining which tags are missing.

The second one is called missing tag identification. The objective is to identify all the missing tags with the required accuracy. This problem is intuitively more difficult than the first one. A straightforward scheme to solve it is to interrogate all the tags one by one~\cite{LiT_2010}. This approach is time-consuming because at least $M=96$ bits need to be transmitted to interrogate each tag. Therefore, a number of subsequent works aimed at improving the time efficiency. Tao \emph{et al.}~\cite{LiT_2010} gave a  lower-bound as $\Theta(N)$ on the average time overhead for outputting all the missing tags, but without formal mathematical proof. They then developed the iterative ID-free (IIP) protocol~\cite{LiT_2010} and the two-hashing (THP) protocol~\cite{LiT_2013}. The former avoids the transmission of tag IDs, while the latter aims to generate more singleton slots by using two hash functions. Zheng \emph{et al.}~\cite{ZhengY_2013} introduced a physical-layer missing tag identification (P-MTI) scheme that extracts information from the aggregated physical-layer symbols rather than individual tag response. For further enhancement, Liu \emph{et al.} designed a multi-hashing-based (MMTI) algorithm~\cite{LiuX_2014} and a slot filter-based algorithm (SFMTI)~\cite{LiuX_2015}. MMTI leverages multiple hash functions for each tag to verify their presence. SFMTI tries to improve the slot utilization by reconciling the expected collision slots as singleton slots. Moreover, Shao \emph{et al.}~\cite{ShaoC_2015} proposed a probabilistic tag retardation (RproTaR) algorithm that exploits a bit vector to eliminate the tag collision and further developed the compact tag transmission. Zhang \emph{et al.}~\cite{ZhangL_2017} proposed a pair-wise collision-resolving (PCMTI) protocol by applying the pair-reply strategy.

Compared with existing works in the literature, we revisit the missing tag identification by quantifying the existing missing tag identification algorithms. 
We then establish a mathematically backed-up lower-bound on the performance of any missing tag identification algorithm. We also developed a missing tag identification algorithm outperforming any state-of-the-art solution a factor of up to $\log N$. Our algorithm employs a novel compact data structure that has not been used in the literature.

\section{Conclusion}
\label{IX}

In this paper, we have revisited the problem of missing tag identification in RFID networks. We have quantitatively compared the existing propositions by showing that the expected execution time of the best solution in the literature is $\Theta \left(N+\frac{(1-\alpha)^2(1-\delta)^2}{ \epsilon^2}\right)$. We have analytically established the expected execution time lower-bound for any missing tag identification algorithm as $\Theta\left(\frac{N}{\log N}+\frac{(1-\delta)^2(1-\alpha)^2}{\epsilon^2 \log \frac{(1-\delta)(1-\alpha)}{\epsilon}}\right)$. Armed with the theoretical result, we have then developed a novel missing tag identification algorithm
by leveraging a tree-based structure with the expected execution time of $\Theta \left(\frac{\log\log N}{\log N}N+\frac{(1-\alpha)^2(1-\delta)^2}{ \epsilon^2}\right)$, reducing the time overhead by a factor of up to $\log N$ over the best algorithm in the literature. Nevertheless, compared to the theoretical performance limit we derive, there is still a performance gap between our algorithm and the theoretical performance limit. Reducing or even filling this gap is already on our research agenda.

\section*{Appendix A \\ Proof of Lemma~\ref{lem:1}}

\begin{IEEEproof}
We prove Lemma~\ref{lem:1} by contradiction. Here we consider a binary version of the \emph{Hamming Distance Estimation} (HDE) problem, namely \emph{Gap-Hamming-Distance} (GHD). A well-known theoretical result proved in~\cite{Chakrabarti_2010} states that solving the GHD problem for any randomized protocol requires at least $\Theta({m})$ bits, given the size of the binary input $m$. Armed with this result,~\cite{ChenB_2013} further proves the following lemma. 

\begin{lemma}
\label{lem:4}
No protocol can solve the HDE problem with $o({m})$ bits of communication while given the size of the binary input $m$, for $\hat{\delta}=1/3$ and $\epsilon<2/{\sqrt{m}}$.
\end{lemma}

 Following Lemma~\ref{lem:4}, we formulate a HDE problem with $\hat{\delta}=\frac{1}{3}>\delta$ and $m=\lceil {\frac{(1-\delta)^2(1-\alpha)^2}{\epsilon^2}} \rceil $, which leads to $\epsilon<\frac{2}{\sqrt{m}}$. Besides, the assumption of $\epsilon \in [\frac{(1-\delta)(1-\alpha)}{\sqrt{N}},1]$ in Lemma~\ref{lem:1} results in $ N\ge m$. Hence, Lemma~\ref{lem:4} can be applied here. 

 Given $L$ slots, Alice and Bob need to virtually execute $\mathcal{P}$ on these slots for $m$ missing identification problem inputs and then count the total number of the missing tags. Suppose that Alice and Bob give $m$ bits of binary strings $x[1],x[2],\dots,x[m]$ and $y[1],y[2],\dots,y[m]$, respectively. To make the missing tag count equal to the Hamming distance between the two strings, tag $i$ could be considered as a present one if $x[i]=y[i]$, or an absent one otherwise. By this way, we can solve the HDE problem as long as all absent RFID tags have been included in the count. In other words, once $\cal{P}$ correctly estimates the RFID missing tag count and outputs an estimation with a relative error no greater than $\epsilon$, Alice and Bob can also use $\mathcal{P}$ to solve the HDE problem with a relative error no greater than $\epsilon$. 

To successfully execute $\mathcal{P}$ on $m$ missing tag identification problem inputs, Alice and Bob are required to find all empty slots in the process of $\mathcal{P}$. Given a slot $s$, Alice and Bob have to develop a predicate $\mathcal{H}^{f}$ by a random string protocol, and then adopt $\mathcal{H}^{f}$ to determine the set of tags that respond in the slot $s$, denoted as $\{1,2,\dots,i\}$. On the one hand, Alice calculates a short string with $u$ bits for $\{1,2,\dots,i\}$ and transmits it to Bob. On the other hand, Bob simultaneously computers a $u$-bit string for $\{1,2,\dots,i\}$ in the same manner. After comparing the two $u$-bite strings, Bob immediately returns the comparison result to Alice with one-bit response. Given that there is no input collisions in the execution of $\mathcal{P}$ (this assumption will be released in the following derivation), Bob replies to Alice with $1$ if $x[i] \neq y[i]$, or $0$ otherwise. The response of $1$ indicates that the tag $i$ is missing and the slot $S$ is empty. On the contrary, the response of $0$ implies that the tag $i$ is present and the slot $S$ is none-empty. By sequentially executing $\mathcal{P}$ on the remaining $L-1$ slots, Alice and Bob can identify all empty slots and further output all the missing tags one by one.  

Recall the above assumptions, we have that $m$ inputs should be hashed to $L$ distinct outputs within $[0,2^{u}-1]$, which guarantees that $\mathcal{P}$ can be successfully executed on these $L$ slots. To avoid the input collision, we select each hash output in turn and repeat the procedure $L$ times. Then, the probability of the $z$-th selection can be denoted as $\frac{1}{{2^{u}-z+1}}$. Let $\mathcal{T}$ denote the event that $\mathcal{P}$ has been successfully executed on these $L$ slots, we have
\begin{equation}
	{P} (\mathcal{T} ) = \prod\limits_{z = 1}^L {\frac{1}{{{2^u} - z + 1}}}  = \frac{{ \left({2^u} - L \right)!}}{{{2^u}!}}.
\end{equation}
By Stirling's approximation~\cite{Dan_2000}, we have
\begin{multline*}
    	P (\mathcal{T} ) \simeq \frac{ \sqrt{2\pi \left ({2^u} - L \right )} {{\left(\frac{{{2^u} - L}}{e}\right)}^{{2^u} - L}}}
	{{\sqrt { {2^ {\left( u+1 \right)}} \pi }} { \left( \frac{2^u}{e} \right)}^{2^u} }
    \simeq \sqrt {1 - \frac{L}{{{2^u}}}} \\ \frac{{\left({2^u} - L\right)\log \left({2^u} - L\right) + L\log e}}{{{2^u}u}}
    >\sqrt {1 - \frac{L}{{{2^u}}}} \dfrac{{2^u} - L}{2^u}.
\end{multline*}

 Then, we define $\mathcal{E}$ as the event that the HDE problem has been solved with a relative error $\leq \epsilon$, and define $\mathcal{V}$ as the event that $\mathcal{P}$ has been successfully executed with a relative error $\leq \epsilon$. 
 To guarantee the constant probability $\hat{\delta}$, the probability ${P} (\mathcal{E})$ should satisfy the following inequalities: 
\begin{equation}
       {P} (\mathcal{E})\geq {P}(\mathcal{V}|\mathcal{T}) {P}(\mathcal{T})
       > 
      \left(1-\delta \right) \sqrt {1 - \frac{L}{{{2^u}}}} \dfrac{{2^u} - L}{2^u}
       \ge 1-\hat{\delta}.
\end{equation}
With the above inequalities, we obtain 
\begin{equation*}
      u \ge \left\lceil {\log \frac{L}{1-\left(\frac{1-\hat{\delta}}{1-\delta}\right)^{\frac{2}{3}}}} \right\rceil.
\end{equation*}

Besides,~\cite{David_2014} has proved that exchanging only $\mathcal {O} (\log m)$ bits can distinguish two $m$-bit strings in the $2$-player $2$-NEQ problem. When given an $(\epsilon, \delta)$-identification-equality protocol $\mathcal P$ and $u = \left\lceil {\log \frac{L}{1-\left(\frac{1-\hat{\delta}}{1-\delta}\right)^{\frac{2}{3}}}} \right\rceil$, Alice and Bob are able to solve the given HDE problem with $\mathcal{O} (L(u+1)+\log m)=\mathcal{O} (L\log L(1-\delta)^{\frac{2}{3}}-L \log(1-\delta)^{\frac{2}{3}}+\log \frac{(1-\delta)(1-\alpha)}{\epsilon})=\mathcal{O} (L\log L+\log \frac{(1-\delta)(1-\alpha)}{\epsilon})$
bits, for $\epsilon<\frac{2}{\sqrt{m}}$ and $\hat{\delta} \in [0,1)$. Particularly, if an RFID missing tag identification protocol $\cal P$ with $L=o(\frac{(1-\delta)^2(1-\alpha)^2}{\epsilon^2 \log \frac{(1-\alpha)(1-\delta)}{\epsilon}})$, Alice and Bob can invoke it as a subroutine to solve the HDE problem with $o(\frac{(1-\delta)^2 (1-\alpha)^2}{\epsilon^2 \log \frac{(1-\alpha)(1-\delta)}{\epsilon}} \log \frac{(1-\delta)^2(1-\alpha)^2}{\epsilon^2 \log \frac{(1-\alpha)(1-\delta)}{\epsilon}} +\log \frac{(1-\alpha)(1-\delta)}{\epsilon})=o(\frac{(1-\delta)^{2}(1-\alpha)^2}{\epsilon^2})=o(m)$ bits, which is in contradiction with Lemma~\ref{lem:4}. Hence, Lemma~\ref{lem:1} is proved.
\end{IEEEproof}

\section*{Appendix B \\ Proof of The First Part of Theorem~\ref{thm:pcmti} on the Performance Bound as a Function of $\epsilon$, $\delta$ and $\alpha$}
\begin{IEEEproof}
Recall the key design of PCMTI, we denote ${P}_{01}$, ${P}_{10}$, and ${P}_{11}$ as the probabilities that a missing tag is assigned to a slot, a missing tag and a present tag are assigned to a slot, and two missing tags are assigned to a slot, respectively. These probabilities are  
	\begin{eqnarray}
      P_{01} &=& \binom{\alpha N}{1} \frac{1}{f}  \left(1 - \frac{1}{f}\right)^{N - 1} \\
P_{10} &=& \binom{\alpha N}{1} \binom{(1-\alpha) N}{1}   \frac{1}{f^2}  \left(1 - \frac{1}{f}\right)^{N - 2} \\
    P_{11} &=& \binom{\alpha N}{2}    \frac{1}{f^2}  \left(1 - \frac{1}{f}\right)^{N - 2},
	\end{eqnarray}
where $f$ is the frame length and the load factor $\rho=\frac{N}{f}$. Let ${P}_{m}$ denote the probability that an arbitrary missing tag is identified in a time slot, we have
$$
          {P}_{m}= { P_{01}}+{ P_{10}}+{ P_{11}}
      =\alpha  \rho  e^{-{\rho}}  (1+\frac{\rho}{2} ).
$$

We then treat each slot as a single observation, and denote $X$ as a random variable representing the observation from a slot. $X=1$ if there are missing tags identified in the slot and $0$ otherwise. Obviously, $X$ follows the Bernoulli distribution, thus 
\begin{gather}
    	{P}\{X=1\}= {P}_{m}=\alpha  \rho  e^{-{\rho}}  (1+\frac{\rho}{2} ),\\
	{P}\{X=0\}= 1-{P}_{m}=1-\alpha  \rho  e^{-{\rho}}  (1+\frac{\rho}{2} ).
\end{gather}
Let $\mathbb{E}[X]$ and $\sigma(X)$ denote the expectation and variance of $X$ respectively, we have
\begin{gather}
       \mathbb{E}[X]=  \alpha  \rho  e^{-{\rho}}  \left(1+\frac{\rho}{2} \right),\\
       \sigma(X)=\sqrt{\alpha  \rho  e^{-{\rho}}  \left(1+\frac{\rho}{2} \right)\left [1-\alpha  \rho  e^{-{\rho}}  \left(1+\frac{\rho}{2}\right )\right ]}.
\end{gather}

We denote $\overline{X}=\frac{1}{n}\sum\nolimits_{i=1}^{n}X_{i}$ as a random variable that is the average of $n$ independent observations, where $X_{1}\cdots X_{n}$ is a random process and $X_{i}$ is the $i$-th observation of $X$. Moreover, $n$ is also a random variable that represents the total number of observations. Since $\rho$ is configured as a constant in PCMTI, $X_{1}\cdots X_{n}$ are independent and identical distribution (i.i.d.). $X_{i}$ and $n$ are independent of each other. Following the law of large numbers~\cite{Etemadi_1981}, we obtain 
\begin{gather}
       \overline{X}=\mathbb{E}\left[\overline{X} \right]=\mathbb{E}[X]=  \alpha  \rho  e^{-{\rho}}  \left(1+\frac{\rho}{2} \right),
       \label{eq:ex}\\
       \sigma \left(\overline{X} \right)=\frac{\sigma(X)}{\sqrt{n}}
       =\sqrt{\frac{\alpha  \rho  e^{-{\rho}}  \left(1+\frac{\rho}{2} \right) \left(1-\alpha  \rho  e^{-{\rho}}  \left(1+\frac{\rho}{2} \right)\right)}{n}}.
       \label{eq:va}
\end{gather}
Let ${P}_{fp}$ denote the false positive in the protocol, i.e., present tags are mistaken for missing ones. Since ${P}_{fp}$ is a very small probability (e.g.\ $10^{-3}$) in RFID missing tag identification protocols, we have $1-{P}_{fp}\simeq 1$. For a single observation, $\overline{X}$ could be utilized to estimate the random variable $\lambda=\frac{\mid \mathbb{A} \cap \mathbb{B}\mid}{\mid \mathbb{A}\mid}$, which is formalized as $\hat{\lambda}=\overline{X} (1-{P}_{fp})/{\alpha}$, where $\hat{\lambda}$ is the estimation of $\lambda$. The identification accuracy requirement can be rewritten as 
\begin{equation*}
{P}  \{\frac{\mid \mathbb{A} \cap  \mathbb{B}\mid}{\mid \mathbb{A}\mid}\ge1- \epsilon \} ={P} \{\overline{X}\ge \frac{(1-\epsilon)\alpha}{1-{P}_{fp}} \}.
\end{equation*}

To determine a guaranteed confidence interval satisfying the above inequalities, we define a random variable $Z=\frac{\overline{X}-\mu}{\sigma}$, where $\mu=\mathbb{E} [\overline{X} ]=\mathbb{E}[X]$ and $\sigma (Z)= \sigma (\overline{X} )=\frac{\sigma(X)}{\sqrt{n}}$. According to the central limit theorem~\cite{Hoeffding_1985}, we know $Z$ follows asymptotically standard normal distribution. Given a particular error probability $\delta$, a total of $r= \lceil \frac{1}{\alpha}-1 \rceil$ trials for $Z$, and $\epsilon \in [\frac{(1-\alpha )(1-\delta)}{\sqrt{N}},\frac{1}{2}]$, we seek a one-sided confidence interval $C$ that satisfies 
\begin{equation}
{P}\{Z\ge C=\overline{X}-\frac{\sigma(Z)}{\sqrt{r}}t_{\delta}(r-1)\}=1-\delta,
\end{equation}
where $t_{\delta}(r-1)$ is a $t$-distribution parameter and $\frac{\overline{X}-\mu}{\sigma(Z)/\sqrt{r}} \sim t_{\delta}(r-1)$. To investigate the expected overhead of PCMTI under the optimal condition, we only consider the case with $\min(n)$. That is to say, when the protocol guarantees ${P}  \{\frac{\mid \mathbb{A} \cap  \mathbb{B}\mid}{\mid \mathbb{A}\mid}\ge1- \epsilon \} \ge 1-\delta$, we will terminate it and compute $\min(n)$, thus yielding the following constraint:
\begin{equation}
\label{eq:con}
 \frac{(1-\epsilon)\alpha}{1-\hat{P}_{fp}}\ge \overline{X}-\sigma(Z)\frac{t_{\delta}(r-1)}{\sqrt{r}}.
\end{equation}

Importing~\eqref{eq:ex} and~\eqref{eq:va} into~\eqref{eq:con} leads to 
\begin{multline}
      \min(n)=\frac{\left(1-\hat{P}_{fp}\right)^2 \hat{P}_{m} \left(1-\hat{P}_{m}\right) \frac{\left[t_{\delta}(r-1)\right]^2}{r}}{\left[\hat{P}_{m} \left(1-\hat{P}_{fp} \right)-(1-\epsilon)\alpha \right]^2}
      = \\\frac{\left[t_{\delta}(r-1)\right]^2}{r} \left \{\frac{\rho  e^{-{\rho}}  \left(1+\frac{\rho}{2}\right)}  {\alpha \left[ \rho  e^{-{\rho}}  \left(1+\frac{\rho}{2}\right) -(1-\epsilon)\right]^2}  \right.  -\\ \left. \frac{\left[\rho  e^{-{\rho}}  \left(1+\frac{\rho}{2}\right)\right]^2}{\left[ \rho  e^{-{\rho}}  \left(1+\frac{\rho}{2}\right)-(1-\epsilon)\right]^2} \right\} 
      = \frac{\left[t_{\delta}(r-1)\right]^2}{r \epsilon^2} \left(\frac{1}{ \alpha}-1\right)
      = \\ \Theta \left(\frac{(1-\alpha)^2(1-\delta)^2}{\epsilon^2}\right).
\end{multline}

Therefore, with such $n$ observations, PCMTI can guarantee the identification accuracy requirement of ${P}  \{\frac{\mid \mathbb{A} \cap  \mathbb{B}\mid}{\mid \mathbb{A}\mid}\ge1- \epsilon \} \ge 1-\delta$. In the above inequality, the $t$-distribution parameter $t_{\delta}(r)$ is proportional to $1-\delta$, and $\rho$ is configured as a constant in PCMTI. Besides, to minimize $n$, we can seek the optimal value $\hat{\rho}$ for $\rho$ with $r=\Theta(\frac{1-\alpha}{\alpha})$, e.g., $19$ with $\alpha=0.05$. Finally, we draw the conclusion that this protocol needs to consume at least  $n=\Theta (\frac{(1-\alpha)^2(1-\delta)^2}{\epsilon^2})$ overhead while guaranteeing an $(\epsilon,\delta)$ identification quality requirement. Since each tag responds to the reader using 
only $\Theta(1)$-bit information, PCMTI requires $\Theta (\frac{(1-\alpha)^2(1-\delta)^2}{\epsilon^2})$ time overhead to output an $(\epsilon,\delta)$ identification quality, for $\epsilon \in [\frac{(1-\alpha)(1-\delta)}{\sqrt{N}},\frac{1}{2}]$ and $\delta \in [0,\frac{1}{3})$. 
\end{IEEEproof}
\bibliographystyle{ieeetr}
\bibliography{Bibliography}

\begin{thebibliography}{10}

\bibitem{ZhuW_2020}
W.~{Zhu}, X.~{Meng}, X.~{Peng}, J.~{Cao}, and M.~{Raynal}, ``Collisions are preferred: {RFID}-based stocktaking with a high missing rate,'' {\em IEEE Transactions on Mobile Computing}, vol.~19, no.~7, pp.~1544--1554, 2020.

\bibitem{Jihong_2020}
J.~{Yu}, W.~{Gong}, J.~{Liu}, L.~{Chen}, K.~{Wang}, and R.~{Zhang}, ``Missing tag identification in {COTS} {RFID} systems: Bridging the gap between theory and practice,'' {\em IEEE Transactions on Mobile Computing}, vol.~19, no.~1, pp.~130--141, 2020.

\bibitem{LiuX_2014}
X.~{Liu}, K.~{Li}, G.~{Min}, Y.~{Shen}, A.~X. {Liu}, and W.~{Qu}, ``A multiple hashing approach to complete identification of missing {RFID} tags,'' {\em IEEE Transactions on Communications}, vol.~62, no.~3, pp.~1046--1057, 2014.

\bibitem{ChiuC_2008}
C.~C. {Tan}, B.~{Sheng}, and Q.~{Li}, ``How to monitor for missing {RFID} tags,'' in {\em Proc. IEEE ICDCS}, pp.~295--302, 2008.

\bibitem{Shahzad_2016}
M.~{Shahzad} and A.~X. {Liu}, ``Fast and reliable detection and identification of missing {RFID} tags in the wild,'' {\em IEEE/ACM Transactions on Networking}, vol.~24, no.~6, pp.~3770--3784, 2016.

\bibitem{JSu_2020}
J.~{Su}, A.~X. {Liu}, Z.~{Sheng}, and Y.~{Chen}, ``A partitioning approach to {RFID} identification,'' {\em IEEE/ACM Transactions on Networking}, vol.~28, no.~5, pp.~2160--2173, 2020.

\bibitem{LiT_2010}
T.~Li, S.~Chen, and Y.~Ling, ``Identifying the missing tags in a large {RFID} system,'' in {\em Proc. ACM MobiHoc}, p.~1–10, 2010.

\bibitem{LiT_2013}
T.~{Li}, S.~{Chen}, and Y.~{Ling}, ``Efficient protocols for identifying the missing tags in a large {RFID} system,'' {\em IEEE/ACM Transactions on Networking}, vol.~21, no.~6, pp.~1974--1987, 2013.

\bibitem{LiuX_2015}
X.~{Liu}, K.~{Li}, G.~{Min}, Y.~{Shen}, A.~X. {Liu}, and W.~{Qu}, ``Completely pinpointing the missing {RFID} tags in a time-efficient way,'' {\em IEEE Transactions on Computers}, vol.~64, no.~1, pp.~87--96, 2015.

\bibitem{ZhengY_2013}
Y.~{Zheng} and M.~{Li}, ``{P-MTI}: Physical-layer missing tag identification via compressive sensing,'' in {\em Proc. IEEE INFOCOM}, pp.~917--925, 2013.

\bibitem{ShaoC_2015}
C.~{Shao}, T.~{Kim}, J.~{Yu}, J.~{Choi}, and W.~{Lee}, ``{ProTaR}: Probabilistic tag retardation for missing tag identification in large-scale {RFID} systems,'' {\em IEEE Transactions on Industrial Informatics}, vol.~11, no.~2, pp.~513--522, 2015.

\bibitem{ZhangL_2017}
L.~{Zhang}, W.~{Xiang}, I.~{Atkinson}, and X.~{Tang}, ``A time-efficient pair-wise collision-resolving protocol for missing tag identification,'' {\em IEEE Transactions on Communications}, vol.~65, no.~12, pp.~5348--5361, 2017.

\bibitem{Roux_2002}
P.~Roux, ``The {ISO}/{IEC} 14443 cards standard,'' in {\em Automatic Fare Collection New Horizons in Public Transport with Smart Cards}, 2002.

\bibitem{Cook_1971}
S.~A. Cook, ``The complexity of theorem-proving procedures,'' in {\em Proc. ACM STOC}, p.~151–158, 1971.

\bibitem{Chakrabarti_2010}
A.~Chakrabarti and O.~Regev, ``An optimal lower bound on the communication complexity of gap-hamming-distance,'' in {\em Proc. ACM STOC}, 2012.

\bibitem{ChenB_2013}
Z.~Zhou, H.~Yu, and B.~Chen, ``Understanding {RFID} counting protocols,'' {\em IEEE/ACM Transactions on Networking}, vol.~24, no.~1, pp.~312--327, 2016.

\bibitem{Yao_1977}
A.~C. {Yao}, ``Probabilistic computations: Toward a unified measure of complexity,'' in {\em Proc. ACM STOC}, pp.~222--227, 1977.

\bibitem{Etemadi_1981}
N.~Etemadi, ``An elementary proof of the strong law of large numbers,'' {\em Zeitschrift Für Wahrscheinlichkeitstheorie Und Verwandte Gebiete}, vol.~55, no.~1, pp.~119--122, 1981.

\bibitem{Hoeffding_1985}
W.~Hoeffding and H.~Robbins, ``The central limit theorem for dependent random variables,'' {\em Duke Mathematical Journal}, vol.~15, no.~3, pp.~349--356, 1985.

\bibitem{Liu_2019}
J.~Liu, X.~Chen, X.~Liu, X.~Zhang, X.~Wang, and L.~Chen, ``On improving write throughput in commodity {RFID} systems,'' in {\em IEEE INFOCOM 2019}, pp.~1522--1530, 2019.

\bibitem{jihong_2017_2}
J.~{Yu}, L.~{Chen}, R.~{Zhang}, and K.~{Wang}, ``On missing tag detection in multiple-group multiple-region {RFID} systems,'' {\em IEEE Transactions on Mobile Computing}, vol.~16, no.~5, pp.~1371--1381, 2017.

\bibitem{jihong_2017}
J.~{Yu}, L.~{Chen}, R.~{Zhang}, and K.~{Wang}, ``Finding needles in a haystack: Missing tag detection in large {RFID} systems,'' {\em IEEE Transactions on Communications}, vol.~65, no.~5, pp.~2036--2047, 2017.

\bibitem{LuoW_2011}
W.~{Luo}, S.~{Chen}, T.~{Li}, and S.~{Chen}, ``Efficient missing tag detection in {RFID} systems,'' in {\em Proc. IEEE INFOCOM}, pp.~356--360, 2011.

\bibitem{LuoW_2012}
W.~Luo, S.~Chen, T.~Li, and Y.~Qiao, ``Probabilistic missing-tag detection and energy-time tradeoff in large-scale {RFID} systems,'' in {\em Proc. ACM MobiHoc}, p.~95–104, 2012.

\bibitem{Dan_2000}
Dan and Romik, ``Stirling's approximation for n!: the ultimate short proof?,'' {\em American Mathematical Monthly}, 2000.

\bibitem{David_2014}
D.~P. Woodruff and Q.~Zhang, ``An optimal lower bound for distinct elements in the message passing model,'' in {\em Proc. ACM-SIAM SODA}, p.~718–733, 2014.

\end{thebibliography}
\end{document}